\documentclass[usenatbib,useAMS]{mnras}

\usepackage{amsfonts}
\usepackage{amsmath}
\usepackage{wasysym}
\usepackage{graphicx}
\usepackage{longtable}
\usepackage{lscape}
\usepackage{soul}
\usepackage{ulem}

\newcommand{\BiPoS}{\mbox{\textsc{BiPoS1}}}
 
\begin{document}

\label{firstpage}

\title[computer programme for binary population synthesis]{\BiPoS$\ $-- a computer programme for the dynamical processing of the initial binary star population}

\author[Dabringhausen, Marks \& Kroupa] {
J. Dabringhausen$^{1}$ \thanks{E-mail: joerg@sirrah.troja.mff.cuni.cz},
M. Marks$^{2}$ \thanks{astro.michi@yahoo.com} and,
P. Kroupa$^{1,3}$ \thanks{pkroupa@uni-bonn.de} \\
$^{1}$ Astronomicky ustav, Universita Karlova, V Holesovickach 2, 180 00 Prague, Czech Republic \\
$^{2}$ Erzbisch{\"o}fliches Clara-Fey-Gymnasium Bonn-Bad Godesberg, Rheinallee 5, 53173 Bonn, Germany \\
$^{3}$ Helmholtz-Institut f{\"u}r Strahlen- und Kernphysik, Universit{\"a}t Bonn, Nussallee 14-16 53115 Bonn, Germany}

\pagerange{\pageref{firstpage}--\pageref{lastpage}} \pubyear{2021}

\maketitle

\begin{abstract}
The first version of the Binary Population Synthesizer (\BiPoS) is made publicly available. It allows to efficiently calculate binary distribution functions after the dynamical processing of a realistic population of binary stars during the first few Myr in the hosting embedded star cluster. Instead of time-consuming N-body simulations, \BiPoS$\ $uses the stellar dynamical operator $\Omega_{\rm dyn}^{\rho_{\rm ecl}}(\log_{10}(E_{\rm b}),t)$, which determines the fraction of surviving binaries depending on the binding energy of the binaries, $E_{\rm b}$. The $\Omega$-operator depends on the initial star cluster density, $\rho_{\rm ecl}$, as well as the time, $t$, until the residual gas of the star cluster is expelled. \BiPoS$\ $has also a galactic-field mode, in order to synthesize the stellar population of a whole galaxy. At the time of gas expulsion, the dynamical processing of the binary population is assumed to effiently end due to the subsequent expansion of the star cluster. While \BiPoS$\ $has been used previously unpublished, here we demonstrate its use in the modelling of the binary populations in the Orion Nebula Cluster, in OB associations and as an input for simulations of globular clusters.
\end{abstract}


\section{Introduction}
\label{sec:introduction}

Stars do not only come as single stars, but they are often bound to a partner star by their gravity. These gravitationally bound stellar systems are called binaries. If they are not disturbed from the outside, they are gravitationally, or dynamically stable. Also systems of more than two stars can be stable over long periods of time, if they are built up hierarchically. There is in principle no limit to the number of hierarchies such a stellar system can have, but important is that on each hierarchy level, the system can be approximated well as two dominant point masses. Thus, the stellar system can be treated as a binary on each level of hierarchy.

Star clusters on the other hand consist also of more than two stars, but they lack the hierachy on the top level of their build-up. Thus, star clusters show the chaotic behaviour of many-body dynamics instead of the regular behaviour of two-body dynamics of hierarchical multiples. It is thought that most, if not all stars are born in embedded star clusters (e.g. \citealt{Kroupa1995a,Lada2003,Bressert2010,Krause2020}). However, embedded star clusters dissolve, first through the removal of their residual gas and then, if they survive, through processes of energy equipartition. Over time, they all leave single stars, binaries and hierarchical multiples in their galaxies.

Observations of the stars in the Galactic Field (GF), that is the stars that are not part of a star cluster any more, allow to fix the ratio of single stars to binaries and to hierachical binaries in the Solar neighbourhood. It turns out that about half of the centre-of-mass stellar systems are single stars. The hiercharchical binaries on the other hand, that is triples, quadruples, quintuples and so on, are comparatively rare against the binaries (e.g. \citealt{Duquennoy1991,Fischer1992,Halbwachs2003,Raghavan2010,Rastegaev2010}). Thus, to first order, the population of centre-of-mass stellar systems in the field can be assumed to consist only of single stars and binaries, which will be done in this paper.

If all stars are born in embedded star clusters, then the field population of the Milky Way must consist of dissolved star clusters, and thus be an aged and dynamically evolved stellar population. This poses the question what the single, binary and hierarchical multiple content of a newly born stellar population is. A good place to look for almost primeval stellar populations are T-Tauri stars, which are stars with a mass $\apprle 3 \, {\rm M}_{\odot}$ and an age of $\apprle 10^7$years, and therefore have not reached the main sequence yet. They are usually found in or near a star-forming gas cloud, that is currently forming embedded star clusters \citep{Joy1945,Kenyon1995}.

For instance, \citet{Kohler1998} observed the Taurus star forming region and found a multiplicity, that is the number of binaries and hierachical binaries over the total number of systems, of $42.5\pm 4.9$ per cent in the range of apparent separations, $s$, from 0.13 to 13 arcsec. This is 1.9 times the multiplicity of in terms of mass and separation comparable binaries in the Galactic Field of 23.5 per cent \citep{Duquennoy1991,Raghavan2010}. The Scopius-Centaurus OB association has a slightly lower multiplicity than the Taurus star forming region, but still notably higher than in the Galactic field \citep{Kohler2000}. An observation of the $\rho$ Ophiuchi Dark Cloud shows however a multiplicity of $29.1\pm 4.3$ per cent in the range of apparent separations from 0.13 to 6.4 arcsec \citep{Ratzka2005}. This corresponds to 18 to 900 AU at a distance of 140 pc, which is about the distance to the $\rho$ Ophiuchi dark cloud and which is at the same distance as the Taurus star-forming region according to \citet{Ratzka2005}. Thus, the binaries observed in the $\rho$ Ophiuchus dark cloud are comparable to those observed in the Galactic field, but their binary fraction is about 1.2 times as high as the binary fraction in the galactic field.

Binaries in young stellar populations are however not restricted to T-Tauri stars, but encompass young stars (and thus also young star clusters) in general. For instance, the Orion Nebula Cluster (ONC) is younger than 1--2 Myr \citep{Hillenbrand1997} and its binary population with periods between $10^{4.8}$ and $10^{6.5}$ days (170 and 6700 years) is comparable to the Galactic Field \citep{Petr1998}. When observing wide binaries with periods of $10^7$ to $10^{8.1}$ days ($2.7\times 10^4$ to $3.4\times 10^5$ years) in the ONC, a lot less binaries than in the Galactic Field are found \citep{Scally1999}. However, the ONC is also very dense with $>10^4$ stars/pc$^{3}$ \citep{Hillenbrand1997}, which implies significant dynamical evolution through stellar interactions. Thus, a probably more primeval population can be found in less dense star clusters \citep{Kroupa2001b}. Such a cluster is for instance NGC 2024, which has about the same age as the ONC, but is a factor 10 less dense, and has a binary fraction much larger than in the Galactic Field in the period range between $10^{5.7}$ and $10^{7.1}$ days (1400 and $3.4\times 10^4$ years). On the other, the cluster IC348 has similar parameters regarding radius and mass as NGC2024, but is about 3--5 times as old, and has again a binary fraction indistinguishable from the Galactic Field in a period range between $10^{5.7}$ and $10^{7.1}$ days (1400 and $3.4\times 10^4$ years, \citealt{Duchene1999}).

Thus, observations show that stellar populations of all ages contain a sizable fraction of binaries. However, are star clusters born with binaries or do these form later?

By-chance encounters of three single stars, where two stars form a binary and the remaining star carries away the excess energy, would require very high densities of about $10^8 \ {\rm M}_{\odot}/{\rm pc}^3$ \citep{Hut1992} to be efficient. Those densities are not observed in the Local Universe (figure~4 in \citealt{Dabringhausen2008}), but may exist for a short time in the most extreme star bursts \citep{Dabringhausen2010,Jerabkova2017}. However, such extreme starbursts are the exception and cannot be responsible for the binary content of the Galactic Field, even if every single star in an extreme star burst would be turned into a binary.

When taking into account that stars are not point masses, tidal captures may play a role. In tidal captures, some of the orbital energy of two stars is converted into tides in a close encounter, and the stars become a binary because of that. However, \citet{Kroupa1995a} showed with numerical simulations of star clusters made up entirely of single stars that tidal captures are also very inefficient to spawn binaries in them. This is somewhat higher in dense environments like the cores of globular clusters, but with $\approx 10^{-7} {\rm yr}^{-1}$ \citep{Hut1992} still too low to explain the observed abundance of binaries even in young clusters.

Thus, processes that turn single stars into binaries are too inefficient to explain present binary numbers. Therefore, a large number of stars must be born as binaries or hierarchical multiples. The hierarchical multiples can be neglected at the birth of a stellar population. They form anyway in the evolution of binary populations, and may reach values consistent with the values observed today in the Galactic Field \citep{Kroupa1995a}. Thus, \citet{Kroupa1995a} proposed that all stars are born in binaries, which are parts of star clusters. A binary fraction of 100 percent is probably a simpification. However, what can be said is that the initial binary fraction is very high, and indistinguishable from being at 100 percent at birth in all observations \footnote{That all stars are born in star clusters with a binarity of 100 percent is also for this paper taken for granted. Thus, stars are not born into a galactic field in this paper, but released into a galactic field from a star cluster. A field population is thus always an evolved population, and its binarity is lower than the 100 percent binarity at its birth. In other words, galactic fields are \textit{not} the dynamically unevolved upper limits for the binary fractions. See Section~(\ref{sec:field}) for the theory, Section~(\ref{sec:impField}) for the implementation into \BiPoS, and Section~(\ref{sec:small}) for an example.}.

In more detail, \citet{Kroupa1995a}, \citet{Kroupa1995b} and \citet{Kroupa1995c} proceeded as follows.

\citet{Kroupa1995a} starts out with a population of binaries of 100 per cent binaries at birth. They are born in star clusters and have formed with a mass function as in \citet{Kroupa1993}. The mass function is modelled based on the luminosity function by \citet{Wielen1983}, with an extension to very low-mass stars described in Section~(4.2) of \citet{Kroupa1993}. The stars of the binaries are paired at random, because of the lack of evidence for a correlation for low-mass stars treated in \citet{Tout1991} and \citet{Kroupa1993}. Low-mass stars, that is stars below a mass of $1 \, {\rm M}_{\odot}$, are however the majority with the mass function from \citet{Kroupa1993}, and are the only stars discussed in \citet{Kroupa1995a}, \citet{Kroupa1995b} and \citet{Kroupa1995c}. The binaries are assumed to be in statistical equilibrium, and thus have a thermal eccentricity distribution (see \citealt{Heggie1975} for a profound treatise). Finally, the binaries have a flat distribution of semi-major axes between $a_{\rm min}=1.69$ AU and $a_{\rm max}=1690$ AU. This distribution of semimajor axes is equivalent to a flat distribution of periods with $\log_{10}(P_{\rm min}/{\rm days})=3.0$ as the minimum period and $\log_{10}(P_{\rm max}/{\rm days})=7.5$ as the maximum period for a mean system mass of $0.64 \, {\rm M}_{\odot}$. The lower and upper limits to the periods are consistent with the observations for pre-main sequence binaries shown in figure~(1) of \citet{Kroupa1995a}. The flatness of the distribution is however an assumption. This assumption only becomes justifiable, because credible assumptions on binary evolution lead to observable populations of main sequence binaries (see below, or \citealt{Kroupa1995b} and \citealt{Kroupa1995c} for more details).

The evolution of the binaries is separated into two modes, namely internal binary evolution in close binaries through the partner star (eigenevolution, \citealt{Kroupa1995b}) and external binary evolution through interactions with other binaries, and later on also with other single stars (stimulated evolution, \citealt{Kroupa1995a}). Internal binary evolution, or pre-main sequence eigenevolution, transforms originally eccentric orbits into circular orbits and sometimes feeds the secondary star from the mass of the primary star. It is especially effective on pre-main sequence binaries with small semi-major axes or strongly eccentric orbits. In both types of orbits, the stars come close to their partner stars, allowing for efficient mass transfer from one star to the other. The reason why it works especially on pre-main sequence binaries is that pre-main sequence stars have larger radii than main sequence stars of the same mass, and are thus more easily disturbed by their companions. \citet{Zahn1989} estimate that it takes $10^5$ years to circularize pre-main sequence binaries. \citet{Kroupa1995b} therefore assumes (and quantifies) that internal binary evolution, or pre-main sequence eigenevolution, takes place in binaries with short semi-major axes and/or high eccentricities and finishes in the pre-main sequence phase of star clusters. He furthermore distinguishes between the birth population, that is the stellar population with parameters as detailed above, and the initial population, that is the stellar population changed by internal binary evolution, or pre-main sequence internal eigenevolution. The binary evolution of the initial binary population can then be followed with an N-body programme over a longer time-span, as done in \citet{Kroupa1995a} with the N-Body code NBODY5 \citep{Aarseth1999}. As a result of internal and external binary evolution, or pre-main sequence eigenevolution and stimulated evolution, a binary population consistent with the galactic field regarding binary fraction, mass ratio distribution, semi-major axis distribution and eccentricity distribution is obtained after one Gyr. The condition for this to be achieved is that all binaries come from star clusters that had 200 binaries on a half-mass radius of 0.8 pc after internal binary evolution (pre-main sequence eigenevolution), or from star clusters that produce the same binary spectrum after their dissolution. Star clusters that have also 200 binaries, but have a noticeably larger or smaller half-mass radius, produce a different binary spectrum. For these reasons, \citet{Kroupa1995a} calls a star cluster that has initially 200 binaries distributed on a half-mass radius of 0.8 pc the {\it dominant-mode star cluster}, and star clusters that produce the same binary spectrum like the dominant-mode star cluster {\it dynamically equivalent}. \citet{Kroupa1995c} and \citet{Kroupa2001b} showed that also the different binary population of well observed star clusters (that is at that time the Plejades, the Hyades, and within observational limitations also the ONC) can be reproduced with the methods devised in \citet{Kroupa1995a,Kroupa1995b}. \citet{Belloni2017} slightly improved the physics of the internal binary evolution, or pre-main sequence eigenevolution. They showed consistency with the present-day population in globular clusters, which must have formed with the same (universal) birth binary formulation \citep{Leigh2015}. Thus, in short, Kroupa's method of dynamically equivalent star clusters \citep{Kroupa1995a,Kroupa1995b,Kroupa1995c,Belloni2018} can be used to characterize the binary populations known so far.

Figure~3 in \citet{Kroupa1995a} shows that most binaries that dissolve into single stars do so in the first Myr, and after that the binary fraction is nearly constant.

Motivated by this, M. Marks developed the first version of a computer programme, which he called \BiPoS$\ $(\underline{Bi}nary \underline{Po}pulation \underline{S}ynthesizer, version 1). It is implicitely introduced already in \citet{Marks2011a}. In this programme, the binary population goes first through  internal binary evolution, or pre-main sequence eigenevolution (cf. \citealt{Kroupa1995b}). Then, the effect of external binary evolution, or stimulated evolution, is calculated with a stellar dynamical operator, $\Omega_{\rm dyn}^{\rho_{\rm ecl}}(\log_{10}(E_{\rm b}),t)$. This operator was introduced in \citet{Kroupa2002} and gives the survival fraction of binaries as a function of the binding energy of the binaries, $E_{\rm b}$, and the time, $t$, for which binary evolution takes place. It depends on the initial density, $\rho_{\rm ecl}$, as determined by the embedded cluster mass at the birth of the star cluster. This density can equivalently be expressed by the embedded mass of the star cluster, $M_{\rm ecl}$, and its half-mass radius at that time, $r_{\rm h}$, which is the approach chosen in \BiPoS.

The operator $\Omega_{\rm dyn}^{\rho_{\rm ecl}}(\log_{10}(E_{\rm b}),t)$ has been gauged in \citet{Marks2011a} to N-body simulations of a set of star clusters with \textsc{Nbody}6 \citep{Aarseth1999} for times of dynamical evolution of 1,~3, and 5~Myr. \BiPoS$\ $makes use of these fitted values and is therefore much faster than the more exact, but also much more time-consuming calculation with an N-body programme. This is because the wider binaries are destroyed quite quickly, but until that happens, they use much of the computing time of the N-body programme. Hence, \BiPoS$\ $provides a shortcut for simulating the first few Myr of evolution of a star cluster, and simulations may become feasible that would otherwise take too long because of the prominent appearance of wide binaries at the beginning of the simulation. In short, \BiPoS$\ $tells the user, in dependency of some parameters, which binaries to keep for eventual further processing with an N-body programme.

An important test for theories of star formation is that the successful theory needs to reproduce the fraction of binaries in dependency of the mass of the primary stars. Most observations have shown that the fraction of binaries decreases as the mass of the primary star decreases. This appears to be in contradiction to the constancy of the fraction of binaries near 100 percent for all primary-star masses at birth, which is assumed here. The observed correlation however arises naturally from this constancy through dynamical processing, as has been shown explicitly in previous work (figure~7 in \citealt{Marks2011b}; figure~6 in \citealt{Thies2015}; and as a basis of prediction in figure~3 in \citealt{Marks2017}).

The purpose of this paper is to introduce \BiPoS$\ $and how it works in more detail. For this, Section~(\ref{sec:theory}) lays out the fundamental equations with which \BiPoS$\ $works. Section~(\ref{sec:implementation}) describes how the equations from Section~(\ref{sec:theory}) are implemented into \BiPoS, Section~(\ref{sec:working}) deals with actually running the programme from the command line and Section~(\ref{sec:examples}) gives examples for running the programme. Section~(\ref{sec:discussion}) is a discussion of some results and Section~(\ref{sec:summary}) concludes the paper.

\BiPoS$\ $can be downloaded at GitHub under the web address \texttt{https://github.com/JDabringhausen/BiPoS1}.

\section{The theory behind \BiPoS}
\label{sec:theory}

In the following sections, binary distribution functions (BDFs) of some orbital parameter $x$ are used, where $x$ is, for instance, the period, $P$, in days, the semi-major axis, $a$, in AU or the eccentricity $e$. They are defined as 
\begin{equation}
\label{eq:bdf}
 \Phi_x=\frac{d f_{\rm bin}(x)}{dx}=\frac{1}{N_{\rm cms}}\frac{d N_{\rm b}(x)}{dx},
\end{equation}
such that
\begin{equation}
\label{eq:bdf-nb}
 f_{\rm bin}=\int\Phi_x(m_1) \, dx.
\end{equation}
In equations~(\ref{eq:bdf}) and~(\ref{eq:bdf-nb}), $f_{\rm bin}=N_{\rm b}/N_{\rm cms}$ is the fraction of binaries, $N_{\rm b}$ is the number of binaries, $N_{\rm cms}$ the number of centre-of-mass systems (that is singles and binaries), and $m_1$ is the mass of the primary of a binary, i.e. the more massive star.

The parameters in the BDFs are for simplicity assumed to be separable upon the formation of the binaries, that is one parameter in the BDF does not depend on any of the others. The observable correlations of the BDFs later on (for example, binaries with short periods, $P$, have low eccentricities, $e$) are due to the subsequent internal binary evolution, or pre-main sequence eigenevolution, leading to the initial binary population (see \citealt{Kroupa1995b, Marks2011a} or Section~\ref{sec:ibp}).

\subsection{Synthesizing the initial binary population from the birth population}
\label{sec:ibp}

\subsubsection{The binary population at its birth}
\label{sec:ibp-1}

The initial binary population (IBP) stems from a birth binary population. The birth binary population undergoes binary-internal evolution, while being in the formation process, which \citet{Kroupa1995b} termed \textit{pre-main sequence eigenevolution}. The birth binary population is not observable, but is a mathematical model which allows the initial binary population to be calculated. The initial binary population is the population which an observer would construct from an observed very young population of stars, if every star could be traced back to its origin, such that all binary systems were reconstructed to an individual age of 0.1 Myr. This can in principle be done in high-resolution radiation hydrodynamical simulations of star formation, as pioneered by \citet{Bate2012}. Thus, also the initial binary population is a theoretical construct.

The periods of the birth binary population, $P$, are selected from a universal period BDF. How this period BDF follows from physical laws is unknown . However, it must fullfill certain requirements, such as that it allows for the binary periods which are observed, or that the integral over all periods equals 1. \citet{Kroupa1995a} find that the function
\begin{equation}
\label{eq:dist-a}
   \Phi_{\rm P,birth}=\delta \frac{\log_{10}P-\log_{10}P_{\rm min}}{\eta+(\log_{10}P-\log_{10}P_{\rm min})^2}
\end{equation}
with the period generating function
\begin{equation}
\log_{10} P (X)=\log_{10}P_{\rm min}+[\delta(e^{2X/\eta}-1)]^{1/2}
\end{equation}
with $X \in [0,1]$ meets these requirements, while it can easily be integrated. \citet{Kroupa1995b} choose $\delta=2.5$, $\eta=45$ and $\log_{10}P_{\rm min}=1$. $\log_{10}P_{\rm max}=8.43$ follows then from the condition that the integration over the whole period range must be unity. These parameters are also adapted in \BiPoS.

For the finding of Equation~(\ref{eq:dist-a}) and its parameters, only primary stars with masses $\le 1 \, {\rm M}_{\odot}$ were used. More massive stars are likely to show somewhat different period distributions, see for instance equation~(3) in \citet{Oh2015}, and for a review \citet{Moe2017}. Also, massive stars are likely not born with 100 percent binaries (or indinguishably close to it), but have also a substantial fraction of primordial triples \citep{Evans2005}, which is potentially connected with the different period functions. However, 84 percent of the stars have a mass $\le 1 \, {\rm M}_{\odot}$ according to the canonical IMF (see equation~\ref{eq:imf} below) which is used in \BiPoS. Moreover, equation~(\ref{eq:dist-a}) gains credibility because it leads to observable period distributions today, if the conditions that follow in Section~(\ref{sec:ibp-2}) regarding the the internal binary evolution, or pre-main sequence eigenevolution, are met.

The primary and secondary component masses for stars with $m<5 \, {\rm M}_{\odot}$ are selected randomly from the canonical stellar initial mass function \citep{Kroupa2001a},
\begin{equation}
 \xi_{\rm IMF}(m)\propto ka_im^{-\alpha_i}
 \left\{
   \begin{matrix}
     \alpha_1=1.3 & 0.08<m<0.5 \\
     \alpha_2=2.3 & 0.5<m<150
   \end{matrix}
 \right.,
 \label{eq:imf}
\end{equation}
where all masses are in ${\rm M}_{\odot}$, and $k$ and $a_i$ are coefficients, which normalize equation~(\ref{eq:imf}) to unity and ensure continuity. For stars more massive than $5 \, {\rm M}_{\odot}$, secondary masses are selected such that the mass ratio $q=m_2/m_1$ is larger than $0.9$ (i.e. close to Unity). Thus, stars with masses $<5 \, {\rm M}_{\odot}$ follow \textit{random pairing} and stars with masses $>5 \, {\rm M}_{\odot}$ follow \textit{ordered pairing} (see also \citealt{Oh2015}). It is important to note that in the case of ordered pairing masses selected from the IMF are not discarded if they do not fullfil the $q$-criterion. Instead they are saved for later use in order to preserve the shape of the IMF.

Thus, the shape of the stellar initial mass function (IMF) is heavily restricted. In practice however, this means little limitations, because the shape of the IMF in most young star clusters is given by equation~(\ref{eq:imf}), or some equation that is observationally indistingushable from it \citep{Dabringhausen2008,Kroupa2013}. Only the upper mass limit changes in low- to mid-mass clusters (see figure~1 in \citealt{Pflamm2007}, figure~2 in \citealt{Weidner2010}; \citealt{Oh2018}). In consequence, the user may change it also in \BiPoS$\ $by setting any value for the upper mass limit, $\texttt{MHIGH}$, provided that it is higher than the lower mass limit, $\texttt{MLOW}$ (see Section~\ref{sec:GenBinaries}). The galaxy-wide IMFs may have different slopes than their star clusters, but that is covered in \BiPoS$\ $by the IGIMF-theory, according to which a galaxy is built up by many embedded star clusters with different upper mass limits for their stars \citep{Marks2011b}. The upper mass limit for stars depends on the mass of the embedded star cluster, and the distribution of embedded star cluster masses mainly (but not only) depends on the star formation rate of the galaxy \citep{Kroupa2003,Jerabkova2018}. What is not covered by \BiPoS$\ $are IMFs that are flatter than $\alpha_2 = 2.3$ in the high-mass range. Such IMFs may occur in the most massive star clusters \citep{Dabringhausen2012,Marks2012b,Jerabkova2017}, and conseqently also in galaxies with the highest star formation rates \citep{Fontanot2017,Jerabkova2018,Dabringhausen2019}.

The binary eccentricities at birth are selected from a thermal BDF,
\begin{equation}
\label{eq:thermal-dist}
 \Phi_{e,\rm birth}=2e.
\end{equation}
This is the thermal distribution function for eccentricties; see Section~(2.2) in \citet{Heggie1975} for a profound treatise, and e.g. \citet{Kroupa1995a,Kroupa2008} for its applications.

\subsubsection{Internal binary evolution, or pre-main sequence eigenevolution}
\label{sec:ibp-2}

The binary population at birth thereby obtained is then subjected to internal binary evolution, or \textit{pre-main sequence eigenevolution}. Internal binary evolution comprises of two aspects: The circularisation of the of the orbits and mass transfer from the (more massive) donor star to the receptor star. Note that both aspects only become relevant when the two stars that make up a binary come close enough together, that is when the binary is very thight, very eccentric, or both.

The driver of the circularisation are the tides that act on eccentric orbits, because at the pericentre, the stars are more strongly harrassed by the gravity of the other star than at their apocentre. (The pericentre, and apocentre, respectively, are the points on the elliptic orbit of a star where the distance to the other star is at the minimum, and maximum, respectively.) The tides heat up the stars, which radiate the energy away. In effect, the apocentre approaches the pericentre as the stars orbit each other, while the pericentre does almost not change. When pericentre and apocentre are equal, the orbit is circular. The stars then have constantly the maximum distortion they originally had anywhere on their orbit. However, with the reason for the tides on the orbit gone, the orbit is stable.

To calculate the change in eccentricity due to pre-main sequence eigenevolution, or internal binary evolution, first the pericenter distance is calculated where eigenevolution is expected to be significant. It is given by
\begin{equation}
 R_{\rm peri}=(1-e_{\rm birth})P^{2/3}(m_1+m_2)^{1/3},
\end{equation}
in which $e_{\rm birth}$ is the eccentricity at birth, $P$ is the period measured in years and $m_1$ and $m_2$ are the masses of the stars measured in ${\rm M}_{\odot}$. The initial eccentricity may then be calculated from
\begin{equation}
 \ln e_{\rm initial} = -T +\ln e_{\rm birth}, \notag
\end{equation}
where
\begin{equation}
 T = \left(\frac{\lambda R_{\odot}}{R_{\rm peri}}\right)^\chi \notag
\end{equation}
is a measure of the duration of internal binary evolution, or pre-main sequence eigenevolution. The parameters $\lambda=28$ and $\chi=0.75$ for internal binary evolution, or pre-main sequence eigenevolution, measure the length-scale over which internal binary evolution of the orbital elements occurs during the proto-stellar phase, and the `interaction strength' between the two protostars in the binary system, respectively.

For the mass transfer, it is important to note that it affects stars during their pre-main sequence phase, that is when they still gather a significant amount of mass from their surroundings. Thus, mass transfer does not necessarily diminish the mass of one star by the amount the other star grows, but the stars can feed from the matter that is not yet part of any star. In fact, \citet{Bonnell1992} proposed that the less massive star could feed on the larger circum-stellar disk of the more massive star. This process would stop at the lastest when the less massive star reaches the mass of the more massive star. On the other hand, \citet{Kroupa1995b} noted that a model, which kept the mass of the binary stars constant in total, proved unsatisfactory when compared with the observational data for binaries with short periods.

Thus, \citet{Kroupa1995b} adopted a feeding model, which is given by
\begin{equation}
 q_{\rm initial} = q_{\rm birth}+(1-q_{\rm birth}) T^*, \notag
\end{equation}
where
\begin{equation}
 T^* = \left\{
 \begin{matrix}
  T, & T\leq1 \\
  1, & T>1
 \end{matrix}
 \right. . \notag
\end{equation}
The initial mass of the secondary component may then be calculated from
\begin{equation}
 m_{2,\rm initial} = q_{\rm initial} \, m_{1,\rm birth},\notag
\end{equation}
and $m_{1,\rm initial}=m_{1,\rm birth}$ is assumed not to change. This model is also taken in \BiPoS.

From the new parameters, the initial period is calculated according to
\begin{equation}
 P_{\rm initial}=\left(\frac{m_{1,\rm birth}+m_{2,\rm birth}}{m_{1,\rm initial}+m_{2,\rm initial}}\right)^{\frac{1}{2}}\left(\frac{1-e_{\rm birth}}{1-e_{\rm initial}}\right)^{\frac{3}{2}}.\notag
\end{equation}

The semi-major axis, energy and angular momentum may now be calculated from the pre-main sequence eigenevolved masses, eccentricities and periods.

Note that with this implementation of internal binary evolution, or pre-main sequence eigenevolution, the periods can only be shortened. However, binaries with periods of $\log_{10}(P/ {\rm days})<-1$ are not allowed in \BiPoS, but small minority at best anyway. In reality, such binaries would likely merge to single stars. For a comparison of a stellar population before and after internal binary evolution in \BiPoS, see figure~(\ref{birth+init}) in this paper, and for the effects in general, see figure~(2) in \citet{Marks2011a}.

\subsection{Synthesizing a star cluster}
\label{sec:omega}

We assume that the embedded cluster is a result of monolithic collapse of a molecular cloud core, because \citet{Banerjee2014} have shown that the observed very young clusters (ONC, NGC3603, R136) are too smooth, compact and young to allow significant sub-structured initial conditions. That is, initial conditions where the final cluster starts forming from the collapse of sub-clusters are constrained to be compact. The sub-clusters are therefore initially so close that the whole structure is dynamically and morphologically next to identical to the assumed monolithic smooth initial conditions (modelled as a Plummer phase-space distribution function). Essentially, sub-clustered initial conditions take too long to collapse and virialise to be consistent with the observed very young clusters.

To calculate the binding energies of the binaries, $E_{\rm b}$, of an evolved star cluster with initial embedded mass in stars, $M_{\rm ecl}$, and initial half-mass radius, $r_{\rm h}$, at first the initial binary energy distribution, $\Phi_{{\log_{10}(E)},\rm init}$ needs to be constructed. \citet{Marks2011a} do this as specified in Section~(\ref{sec:ibp}), and then they calculate the binding energies of the binaries in the star cluster with
\begin{equation}
 E_{\rm b}=2^{-\frac{1}{3}}\left(\frac{\pi m_1m_2}{P}\right)^{\frac{2}{3}}.\notag
\end{equation}
After that, \citet{Marks2011a} perform N-body simulations, using \textsc{Nbody6} \citep{Aarseth1999,Aarseth2003}, to transform the initial binding energies into the final ones. They use an array of star clusters of different $M_{\rm ecl}$, $r_{\rm h}$ and ages, $t$, for this step. The result is an array of evolved binary energy distribution functions, $\Phi_{\log_{10}(E_{\rm b}),\rm evolved}$, which depend on $M_{\rm ecl}$, $r_{\rm h}$, and the age, $t$, of the star cluster. Alternatively, $M_{\rm ecl}$ and $r_{\rm h}$ can be replaced with a single parameter, $\rho_{\rm ecl}$, that is the average density of the star cluster within $r_{\rm h}$, because it has been demonstrated in \citet{Marks2011a} that star clusters with identical crossing times, $t_{\rm cross}$, develop identical binary fractions, $f_{\rm b}$. And since $t_{\rm cross} \propto \rho_{\rm ecl}^{-0.5}$, the initial density is the only relevant parameter to determine the resulting binary population in their computations.

When the initial binary population (IBP) is placed inside a star cluster, the IBP will evolve due to interactions between systems in which energy and angular momentum is transferred. This is called external binary evolution, or stimulated evolution. Generally, external binary evolution, or stimulated evolution, removes binaries from the population with time, and the amount to which this happens depends on the binding energy of the binary. The change of the IBP due to external binary evolution, or stimulated evolution, until the time $t$ can be written down as a stellar dynamical operator, $\Omega_{\rm dyn}^{M_{\rm ecl},r_{\rm h}}$, which acts on the initial energy BDF. Thus,
\begin{equation}
\label{eq:operator}
 \Phi_{\log_{10}(E_{\rm b}),\rm evolved} = \Omega_{\rm dyn}^{M_{\rm ecl},r_{\rm h}}(\log_{10}(E_{\rm b}),t) \times \Phi_{\log_{10}(E_{\rm b}), \rm initial}.
\end{equation}
$E_{\rm b}$ is the binding energy of the binaries, and the superscipts $M_{\rm ecl}$ and $r_{\rm h}$ on the operator signify its dependence on $M_{\rm ecl}$, and $r_{\rm h}$, respectively, of the star cluster in question. Equivalently, $\Omega_{\rm dyn}^{M_{\rm ecl},r_{\rm h}}(\log_{10}(E_{\rm b}),t)$ can be replaced by $\Omega_{\rm dyn}^{\rho_{\rm ecl}}(\log_{10}(E_{\rm b}),t)$, where $\rho_{\rm ecl}$ is the average initial density of the embedded star cluster, corresponding to the above combination of $M_{\rm ecl}$ and $r_{\rm h}$.

The next step is to characterize $\Omega_{\rm dyn}^{M_{\rm ecl},r_{\rm h}}(\log_{10}(E_{\rm b}),t)$, or $\Omega_{\rm dyn}^{\rho_{\rm ecl}}(\log_{10}(E_{\rm b}),t)$, respectively. \citet{Marks2011a} find that $\Omega_{\rm dyn}^{\rho_{\rm ecl}}(\log_{10}(E_{\rm b}),t)$ can be described as the upper half of a sigmoidal curve, which is given as
\begin{multline}
\label{eq:sigmoidal}
\Omega_{\rm dyn}^{\rho_{\rm ecl}}(\log_{10}(E_{\rm b}),t)=\\
\frac{{\cal A}(t)}{1+\exp[{\cal S}(t)(\log_{10}(E_{\rm b})-\log_{10}(E_{\rm b,cut}))]}-\frac{{\cal A}(t)}{2}.
\end{multline}
The parameters ${\cal A}$, ${\log_{10} (E_{\rm b,cut}})$ and ${\cal S}$ can be interpolated as functions of $\rho_{\rm ecl}$ and $t$. They are given as
\begin{equation}
\label{eq:operator2}
 {\cal A}(t) = \left\{
	    \begin{matrix}
	     a(t)+b(t)\log_{10}\rho_{\rm ecl} & \mbox{if result}>-3.2 \\
	     -3.2 & \mbox{otherwise}
	    \end{matrix}
	    \right.,
\end{equation}
\begin{equation}
\label{eq:operator3}
 \log_{10}(E_{\rm b,cut}) = \left\{
	    \begin{matrix}
	     c(t)+d(t)\log_{10}\rho_{\rm ecl} & \mbox{if result}\leq2 \\
	     2 & \mbox{otherwise}
	    \end{matrix}
	    \right.,
\end{equation}
and
\begin{equation}
\label{eq:operator4}
 {\cal S}(t) = -\frac{1}{\exp[e(t) \times (\log_{10}\rho_{\rm ecl}-f(t))]}-g(t).
\end{equation}
The time-dependent coefficients $a(t)$ to $g(t)$ in equations~(\ref{eq:operator2}) to~(\ref{eq:operator4}) are listed in table~(2) in \citet{Marks2011a} for the ages of of~1,~3, and $5\, {\rm Myr}$. These ages are also the available choices for \BiPoS. They can be interpreted in this context as the times for which external binary evolution, or stimulated evolution, acts on the star clusters. Note that the simulations on which $\Omega_{\rm dyn}^{\rho_{\rm ecl}}(\log_{10}(E_{\rm b}),t)$ is gauged do not include gas expulsion. The expansion of the star clusters, which inhibits binary-binary interactions, and later on also binary-single interactions, is caused by the energy that is set free by these processes themselves. Thus, $\Omega_{\rm dyn}^{\rho_{\rm ecl}}(\log_{10}(E_{\rm b}),t)$ saturates, and after $5 \, {\rm Myr}$ at the latest, star clusters have expanded so much that encounters between binaries happen only rarely. However, also gas expulsion will happen in the first few Myr, and also lead to an expansion or even a dissolution of the star clusters. The binary populations of such star clusters do not evolve much beyond this point in time, but become `frozen in'.

In principle, binaries of a certain binding energy can still have the full range of eccentricities, $e$, between 0 (circular orbit) and 1 (radial orbit). Radial orbits with a certain binding energy, $E_{\rm b}$, are easier to destroy by encounters with other stars than circular orbits with the same $E_{\rm b}$. The reason is that the radial orbits are more lightly bound than the average value of $E_{\rm b}$ for most of the orbital period spent at larger distances. The circular orbits are, in contrast, always bound with the average $E_{\rm b}$. In practice however, the effect of different eccentricties is secondary next to the effect of different $E_{\rm b}$, see figure~8 in \citet{Marks2011a}. Also that internal binary evolution, or pre-main sequence eigenevolution (see Section~\ref{sec:ibp}), is making eccentric orbits more circular diminishes the problem. High eccentricities remain after internal binary evolution, or pre-main sequence eigenevolution, only for the wide binaries, which are weakly bound and are therefore easy to destroy with external binary evolution, or stimulated evolution, independent of their $e$. Thus, only considering $E_{\rm b}$ proves to be sufficient for the purpose of \BiPoS.

\BiPoS$\ $could also be adapted for highly substructured star clusters, and not just for the monolithic case considered here. For this, the coefficients $a(t)$ to $g(t)$ obtained by the fits of equations~(\ref{eq:operator2}) to~(\ref{eq:operator4}) would potentially be different, but the general code of \BiPoS$\ $would remain the same. Note however that \citet{Parker2011} run computations of both substructured (clumpy) and rather spherical star cluster setups with initially 100 percent binaries and different binary distribution functions. Among them is also the IBP by \citet{Kroupa1995b}. They find that the resulting binary fractions are a weak function of `clumpiness': Substructured clusters produce up to $\approx$10 percent lower binary fractions after 10 Myr of dynamical evolution when compared to more spherical setups. However, a population initially dominated by binaries is processed strongly in both types of clusters. While in a spherical setup, the processing of binaries depends on their density in their core, the driver of breaking up the binaries is in clumpy clusters the density in the substructures \citep{Parker2011}. Although \BiPoS$\ $has been gauged in simulations from spherical cluster setups, this finding allows to interpret the cluster density passed to \BiPoS$\ $as the density in such clumps in the substructure of a star cluster. The clumps in substructured star cluster later merge, e.g. after a cool collapse, to form the cluster population seen nowadays.

\subsection{Synthesizing a galactic field}
\label{sec:field}

Adding up the stellar populations from dynamically evolved and eventually dispersed star clusters yields a galactic field population. Young star clusters follow an embedded cluster mass function (ECMF) described by power-law index $\beta$,
\begin{equation}
\label{eq:ECMF}
 \xi_{\rm ECMF}(M_{\rm ecl}) \propto M_{\rm ecl}^{-\beta}.
\end{equation}
An integrated galaxy-wide field BDF (IGBDF) is then arrived at evaluating
\begin{equation}
\label{eq:IGIMF}
 \Phi_{x}^{\rm field} = \int_{M_{\rm ecl,min}}^{M_{\rm ecl,max}({\rm SFR})} \Phi_{x,\rm evolved}^{\rm cluster} \xi_{\rm ECMF}(M_{\rm ecl}) d M_{\rm ecl},
\end{equation}
where the $x$ stands for the observed orbital parameter (for instance $q$, $e$, and so on). The limits $M_{\rm ecl,min}$ and $M_{\rm ecl,max}({\rm SFR})$ are the mass of the star cluster with the lowest stellar mass and star cluster with the highest stellar mass, respectively, in the star cluster system (SCS). The maximum mass $M_{\rm ecl,max}$ depends on the star formation rate (SFR) with which the SCS has formed and is calculated from
\begin{equation}
\label{eq:SFR-clustermass}
 \frac{M_{\rm ecl,max}}{{\rm M}_{\odot}}=84793\times\left(\frac{{\rm SFR}}{{\rm M}_{\odot} \, yr^{-1}}\right)^{0.75}
\end{equation}
according to \citet{Weidner2004b}.

Each cluster selected from the above ECMF will contribute its own final binary population depending on its initial density, where lower-mass clusters, on average, will retain a larger binary population, which is contributed to the field population. The individual binary populations that contribute to the field population of a galaxy are found for each star cluster following the recipe in Section~\ref{sec:omega}.

\section{Implementation in \BiPoS}
\label{sec:implementation}

\subsection{Generating a library of binaries}
\label{sec:GenBinaries}

Before calculating the final properties of a population of binaries, \BiPoS$\ $needs to create a birth binary population and calculate the initial binary population from it. The shape of the initial stellar mass function (IMF) is set to the canonical IMF (\citealt{Kroupa2001a} and equation~\ref{eq:imf} in this paper), but the user can choose the lower and upper mass limit of the IMF, $\texttt{MLOW}$ and $\texttt{MHIGH}$, as well as the total number of binaries to be generated, $\texttt{Nlib}$. The part of the programme responsible for creating the initial binary population is archived in the file \texttt{Library.c}

For creating the birth binary population, \BiPoS$\ $chooses $2 \times \texttt{Nlib}$ stellar masses, that are randomly selected from the IMF (eq.~\ref{eq:imf}), and stores them into an array. The IMF is interpreted here as a pure probabilistic function in the mass interval $[\texttt{MLOW} \ge 0.08 \ {\rm M}_{\odot}, \texttt {MHIGH} \le 150 \, {\rm M}_{\odot}]$, where $\texttt{MLOW} < \texttt{MHIGH}$ \footnote{The IMF can also be interpreted as an \textit{optimal distribution function} \citep{Kroupa2013}. In this case, stars can be paired randomly from an array of optimally sampled masses.}. Thus, together with the canonical IMF (equation \ref{eq:imf}), the normalisation condition
\begin{equation}
1=k a_i\int_{\texttt{MLOW}}^{\texttt{MHIGH}} (m')^{-\alpha_i} \, dm'
\label{eq:imfnorm}
\end{equation}
is used to determine the coefficients such that the IMF is continuous. The constant $k$ is a normalisation that guarantees that, together with the choices for the $a_i$, the integration of the right side of equation~(\ref{eq:imf}) equals 1.

In order to select a mass from the IMF determined from equation~(\ref{eq:imf}) and~(\ref{eq:imfnorm}), the cumulative initial mass distribution is mapped to a uniform random variate $X(m)\in[0:1]$, such that
\begin{equation}
\label{eq:randomIMF}
 X(m) = ka_i\int_{\texttt{MLOW}}^{m}(m')^{-\alpha_i} \, dm' .
\end{equation}
Thus, by integrating equation~(\ref{eq:randomIMF}) and solving it for $m$, the stellar mass $m$ in dependency of the uniform random variate $X(m)$ is obtained. By doing this for the $X(m)$ values, the distribution of the $2 \, \texttt{Nlib}$ values for $m$ is consistent with the canonical IMF.

\BiPoS$\ $then pairs the stars to binaries by looking at the mass of the next star at first in the array. If the star is less massive than $5 \, {\rm M}_{\odot}$, it is simply paired with the following star in the list that is also less massive than $5 \, {\rm M}_{\odot}$, while stars more massive than $5 \, {\rm M}_{\odot}$ are overlooked. This produces random pairing of the stars less massive than $5 \, {\rm M}_{\odot}$, because the stars in the array are not sorted. If the star is more massive than $5 \, {\rm M}_{\odot}$, then the programme searches for the next star that together with the first star has a mass ratio $q>0.9$. Thus, stars more massive than $5 \, {\rm M}_{\odot}$ usually have partner stars of almost the same mass, which corresponds to ordered pairing. When for a star with a mass larger than $5 \, {\rm M}_{\odot}$ no companion star is found such that $q>0.9$, then simply the next star in the array is picked, independent of its mass, as in the random pairing procedure. This is necessary as one cannot simply add more massive stars to fullfill the mass-ratio criterion since this would change the underlying IMF. However, the effect of random pairing of a few massive stars becomes negligible for a large $\texttt{Nlib}$, because almost every star with a mass $>5 \, {\rm M}_{\odot}$ finds a suitable partner star in that case. The programme goes through the array of stars until every star is part of a binary.

\BiPoS$\ $then assigns the period $P$ using equation~(\ref{eq:dist-a}). The programme proceeds analoguous to the creation of stars from the IMF, that is the inverse of the integral of equation~(\ref{eq:dist-a}) is formed, and then uniform random variates between 0 and 1 are mapped onto this function. A random variate $X \in [0,1]$ thereby produces the correct distribution in $P$ (see chapter 7.2 in \citealt{Press1992} for a more profound treatise of the topic). The eccetricities $e$, which are at birth distributed according to equation~(\ref{eq:thermal-dist}), are assigned here analogously.

By default, the birth values for $e$, $P$, $m_1$ and $m_2$ are then subjected to internal binary evolution, or pre-main sequence eigenevolution (see \citealt{Kroupa1995b} or Section\ref{sec:ibp}) to arrive at the initial values.

The logarithmic binding energies $\log_{10}(E_b)$ (in $\log_{10}[{\rm M}_{\odot} \, {\rm km}^2 \, {\rm s}^{-2}]$), the logarithmic angular momenta $\log_{10}(L)$ (in $\log_{10}[\rm cm^2 \, s^{-1}]$), and the logarithmic semi-major axes $\log_{10}(a)$ (in $\log_{10}[\rm AU]$) follow from $P$, $m_1$ and $m_2$.

The apparent separations $s$ are projections on the sky of the correspondent $a$ that are obtained as follows. There is a vector $\mathbf{r}=(r,\phi,\cos(\theta))$ that determines the orientation of each binary with respect to the observer. The component of the vector determining the radius $r$ is set fixed to 1, but for the angles $\phi$ and $\theta$, random variates are chosen, so that the values for $\phi$ are distributed uniformly between 0 and 2$\pi$, and the values for $\theta$ are distributed uniformly between $-\pi$ and $\pi$. Thus, the semimajor axis rotates with $\phi$ between 0 and 2$\pi$, and $\theta$ determines whether the observer sees the binary face-on ($\cos(\theta)=\pm 1$), egde-on ($\cos(\theta)=0$), or somewhere in between. The projected separations are obtained by projecting the 3-dimensional vectors $\mathbf{r}$ onto an arbitrary, but then fixed plane; for instance the $x-y$-plane.

Also the quantities $t/{\cal T}$ and $\varphi$ are listed in the output file. These parameters tell the user the orbital-time and phase in which the binary can currently be found, and are distributed uniformly between 0 and 1, and 0 and 2$\pi$, respectively.

About internal binary evolution, or pre-main sequence eigenevolution, note that it can in principle lower the fraction of binaries born in a star cluster. For this, the orbits become so tight that the stars merge to single stars. However, this happens rarely, as can be seen in Section~(\ref{sec:birth-initial}). Thus, the binary fraction remains here where it is during internal binary evolution, or pre-main sequence eigenevolution; that is at 100 percent. The only way to lower this value is external binary evolution, or stimulated evolution, which will be dealt with in Sections~(\ref{sec:impCluster}) and~(\ref{sec:impField}).

The user may, besides making choices for $\texttt{MLOW}$, $\texttt{MHIGH}$ and $\texttt{Nlib}$, also turn off pre-main sequence eigenevolution and ordered pairing for stars more massive than $5 \, {\rm M}_{\odot}$ in the creation of the file of initial binary properties. The programme then delivers the birth population of binaries (pre-main sequence eigenevolution off) and pairs stars over the whole mass range completely at random to binaries (ordered pairing off) to test their influence on the resulting populations. However, this is not recommended in the light of the literature concerning pre-main sequence eigenevolution (see \citealt{Kroupa1995b, Belloni2017}) and ordered pairing of massive binaries (see e.g. \citealt{Kobulnicky2007,Sana2008,Sana2009}). This should therefore only be done for comparisons to the more realistic populations, where eigenevolution and ordered pairing for massive stars are turned on (the default).

\subsection{Implementing external binary evolution, or stimulated evolution}
\label{sec:impExt}

\subsubsection{General remarks}
\label{sec:impGenRemarks}

The user has now two possibilities to proceed when \BiPoS$\ $has created an initial stellar population from a birth stellar population, namely by assuming that the initial population is a single star cluster (see Section~\ref{sec:impCluster}), or by assuming that the initial population is the basis for a field population made up from multiple dissolved embedded star clusters (see Section~\ref{sec:impField}). However, some general remarks first.

Generally, each encounter of a binary either destroys the binary, or it transforms its set of initial binary parameters into a new set. Thus, in reality, also the encounters that leave the binary intact would change its orbital parameters. \BiPoS$\ $however concentrates only on how many binaries per energy bin are destroyed. Figures~(8) and~(10) in \citet{Marks2011a} show that the data obtained with N-body simulations agree very well with those obtained analytically (that is with \BiPoS). This gives confidence that the approximation, that \BiPoS$\ $makes, is valid. Thus, if \BiPoS$\ $lets a binary survive, it does so with all initial binary parameters unchanged.

If the user decides to use \BiPoS$\ $in the star cluster mode, the user can specify a cluster mass from the command line. This might seem like a double effort at first, since the user has already set up a library of binaries, corresponding of a specific total mass. However, the purpose of the library is predominantely to control statistical uncertainties occuring especially when dealing with small star clusters. Thus, if the user is interested in the properties of an \textit{average} small star cluster, the user is still encouraged to set the size of the library to $\apprge 10^7$ binaries (also the default in \BiPoS), even though the total mass of the star cluster will be much smaller. The user should instead choose $\texttt{MLOW}$ and $\texttt{MHIGH}$ according to the problem to be covered. For instance, if the user decides to investigate very small star clusters of $10 \, {\rm M}_{\odot}$, the library should definetely not hold stars more massive than $10 \, {\rm M}_{\odot}$, and according to \citet{Weidner2010} the actual highest stellar mass is much lower still. However, the programme would run in any case smoothly through the whole process, also when the library contains stars with masses up to $150 \, {\rm M}_{\odot}$. The mass of the star cluster does, however, matter for its evolution. This can be seen for instance by comparing star clusters of the same half-mass radius, the same evolution time and the same library of binaries, but with different masses. The less massive star clusters will then have more surviving binaries because of their, on average, lower densities. Thus, ultimately the fraction of surviving binaries decreases with increasing $\rho_{\rm ecl}$.

The part of \BiPoS, which is responsible for external binary evolution, or stimulated evolution, is archived in the file \texttt{Synth.c}. As the programme proceeds, \texttt{Synth.c} calls a number of functions, which are archived in \texttt{InitFinDistr.c}.

\BiPoS$\ $finally stores the requested information into table(s) in the folder \texttt{output/}; that is the requested, binned binary parameters of the surviving binaries. Note that even though only the energy distribution is considered when running \BiPoS, the output files will contain the requested orbital parameters, i.e. also different from $E_{\rm b}$ (see Section~\ref{sec:impCluster}). Also the default number of bins for the requested parameter(s) may be (chosen) different from the default for energy bins.

All these output tables have three columns. The first column contains the centres of the bins of the chosen orbital parameter. The second column holds the normalised binary fractions, that is the binary fraction divided by the bin width. The third column stores the absolute number of binaries per bin, if in an observation \texttt{N} targets have been searched for multiplicity. This number results from a multiplication of the second column with \texttt{N} and the bin width. The default value for \texttt{N} is 100; a different value may be set by the user using the command-line (see Section~\ref{sec:command-line}). Using the default value, the last column will hold the percentage of the stars which are part of a binary, that is the binary fraction in percent.

\subsubsection{Implementing star clusters}
\label{sec:impCluster}

The function \texttt{Synthesize(...)} in the file \texttt{Synth.c} performs the tasks required to synthesize a single star cluster population (this section) or the population in a whole galactic field (Section~\ref{sec:impField}).

It starts by reading the textfile \texttt{flE\_eigenevolved.dat}, which is included in the \BiPoS$\ $ package. The textfile \texttt{flE\_eigenevolved.dat} contains the initial energy distribution, that is the one after internal binary evolution, or pre-main-sequence eigenevolution. It lists binary fractions of $\log_{10}(E_{\rm b})$ into an array named \texttt{ledf[]}. The small logarithmic energy-intervals in \texttt{flE\_eigenevolved.dat} range from $\log_{10}(E_{\rm b})=-6.0$ to $\log_{10}(E_{\rm b})=6.0$ in steps of $0.01$ and in units of $\log_{10}[{\rm M}_{\odot} \, {\rm km}^2 \, {\rm s}^{-2}]$. They are read into the array and normalised to the width of the equidistant bins through a call of the function \texttt{edf(...)} located in the file \texttt{InitFinDistr.c}. The array \texttt{ledf[]} then contains the initial energy distribution on which the stellar dynamical operator acts (eq.~\ref{eq:operator}).

\BiPoS$\ $then retrieves the values of the parameters ${\cal A}$, ${\log_{10} (E_{\rm b,cut}})$ and ${\cal S}$, which enter the stellar dynamical operator $\Omega_{\rm dyn}^{\rho_{\rm ecl}}(\log_{10}(E_{\rm b}),t)$ (eq.~\ref{eq:sigmoidal}) using eqs.~(\ref{eq:operator2}),~(\ref{eq:operator3})~and~(\ref{eq:operator4}). These equations result directly from the time, $t$, for stimulated evolution and the initial cluster density within the half-mass radius, $\rho_{ecl}$, provided by the user through the command-line. The $\Omega_{\rm dyn}^{\rho_{\rm ecl}}(\log_{10}(E_{\rm b}),t)$ operator can now be evaluated and its function values are stored in another array named \texttt{sigm[]} of the same size as the energy array generated previously.

Through a simple element-by-element multiplication of the \texttt{sigm[]} and \texttt{ledf[]} arrays (cf. eq.~\ref{eq:operator}), the final energy array \texttt{res[]} arises as a result, which contains the normalised binary fractions after external binary evolution, or stimulated evolution. In the function \texttt{distribute\_nrgs(...)}, the final distribution is then transformed into an energy array \texttt{nrgspect\_fin[]}. The array \texttt{nrgspect\_fin[]} contains the \textit{number} of binaries in each bin, through a multiplication of each element in \texttt{res[]} with the binwidth and the final number of centre-of-mass systems, that is the added numbers of singles and binaries after stimulated evolution. A similar array \texttt{nrgspect\_in[]}, which holds the numbers of initial binaries per bin, is created for the initial energy distribution contained in \texttt{ledf[]} and the initial number of centre-of-mass systems.

The initial number of systems is simply half the number of stars selected from an IMF in a population of 100\% binaries. The final number of systems is calculated from the final binary fraction. This final binary fraction is the sum of the entries in \texttt{res[]} multiplied by the binwidth, and the average number of stars in the considered cluster. The latter is calculated from the division of the initial cluster mass, $M_{ecl}$, and the average mass of the canonical IMF.\footnote{For the cluster mode in \BiPoS$\ $the number of systems used to calculate the number of binaries per energy bin can in principle be any number and does not need to be calculated from the initial cluster mass, since \texttt{nrgspect\_fin[]} is scaled in the next step, anyway. Here this procedure has been chosen since the relative numbers of stars in clusters of different mass become important when \BiPoS$\ $is run in the field-mode (Section~\ref{sec:impField}).}

The array \texttt{nrgspect\_fin[]} contains no information whatsoever about other orbital parameters distributions. In order to extract other parameters from the the resulting energy distribution for binaries, the library of binaries created before, as described in Section~(\ref{sec:GenBinaries}), comes in. Calling the function \texttt{populate\_initial\_energy\_distribution(...)}, \texttt{Synthesize(...)} has read the user-generated library at its start into an array \texttt{fE\_in[]}, which is coded in the file \texttt{InitFinDistr.c}. The function \texttt{Synthesize(...)} passes this array alongside \texttt{nrgspect\_in[]} and \texttt{nrgspect\_fin[]} to the function \texttt{orbital\_parameter\_distributions()} contained in the same file, in order to extract different orbital parameters.

The idea is to break up dissolved binaries in the user-generated library into its constituents and retain only a number of binaries in this library that correspond to the energy distribution after external binary evolution, or stimulated evolution. Information about the retained binaries is contained in \texttt{nrgspect\_fin[]}. Since the number of binaries in \texttt{nrgspect\_in[]}, however, does in no way resemble the number of binaries in the user-generated library contained in \texttt{fE\_in[]}, each entry, $i$, in the final energy array \texttt{nrgspect\_fin[]} is scaled by the factor \texttt{fE\_in[$i$]/nrgspect\_in[$i$]}.

The programme then reads the user-generated library (again) and loops through each of its entries. The binding energy of the binary is compared to \texttt{nrgspect\_fin[]} until the corresponding energy bin has been found. Is the number of binaries in this bin larger than 0, the binary in the library is left intact and the number of binaries in the \texttt{nrgspect\_fin[]} bin is reduced by 1. Does the number of binaries in this bin equal 0, all binaries in the final energy spectrum after stimulated evolution have been distributed and the binary in the library is broken up. Dissolved binaries are counted as two single stars and two centre-of-mass systems, while an intact binary counts as one binary and one centre-of-mass system. This procedure is continued until all entries in the \texttt{nrgspect\_fin[]} array contain zeroes without exception. Thus, \BiPoS$\ $leaves the first binaries in the library intact and destroys those coming last in the library. However, for a sufficiently large library, the biases concerning surviving and destroyed binaries are negligible.

For each orbital parameter in the library of binaries an individual array has been created. If a binary did not dissolve during the procedure before, a binary is added (+1) in the corresponding bin of the respective orbital parameter array. This way distributions for all other orbital parameters for surviving binaries can be extracted.

\BiPoS$\ $finally stores the requested orbital parameter distributions into table(s) in the folder \texttt{output} (see the end of Section~\ref{sec:impGenRemarks}).
  
\subsubsection{Implementing galactic fields}
\label{sec:impField}

\BiPoS$\ $works for galactic fields in principle the same as for star clusters. The only difference is that for galactic fields, \BiPoS$\ $additionally deals with an embedded star cluster mass function (ECMF) instead of a single star cluster. To handle this problem, \BiPoS$\ $assumes that a star cluster population (SCP) is fully populated after a certain time $\delta t$. Thus,
\begin{equation}
\label{eq:deltat}
M_{\rm scp}=\delta t \times SFR,
\end{equation}
where $M_{\rm scp}$ is the mass of the star cluster population and the star formation rate (SFR) is set by the user . \citet{Weidner2004b} found a universal value for $\delta t$, which they give as $\approx 10$ Myr. This is the timescale over which the interstellar medium of a galaxy forms a new population of embedded star clusters of combined stellar mass $M_{\rm scp}$ (see also \citealt{Schulz2015}). Therefore, $\delta t = 10$ Myr is also set as a fixed value for \BiPoS. Thus, for example, a galaxy like the Milky Way with a SFR of $3 \, {\rm M}_{\odot} \, {\rm yr}^{-1}$ (consistent with \citealt{Prantzos1995}, and the default for the SFR in \BiPoS) has always $M_{\rm scp}=3\times 10^7 \, {\rm M}_{\odot}$, independent of the changeable parameters in \BiPoS. 

One of these parameters the user may change is the exponent $\beta$ of the embedded cluster mass function (ECMF; e.g. \citealt{Elmegreen1997,Lada2003} or equation~\ref{eq:ECMF} in this paper). The default value for $\beta$ set by \BiPoS$\ $is $\beta=2$, as proposed by \citet{Elmegreen1997} and \citet{Lada2003}. The ECMF is needed to estimate its normalisation factor $k_{\rm ecl}$. For this, using
\begin{equation}
\label{eq:normalisation}
M_{\rm scp}=k_{\rm ecl} \int_{M_{\rm ecl,min}}^{M_{\rm ecl,max}(SFR)} M_{\rm ecl}^{- \beta} dM_{\rm ecl}
\end{equation}
is used, where $M_{\rm ecl}$ is the stellar mass of the embedded star clusters, $M_{\rm ecl,min}$ is the lowest mass for embedded star clusters and $M_{\rm ecl,max}(SFR)$ the highest mass for embedded star clusters. Thus, the normalisation also depends on $\beta$. With higher values for $\beta$, the ECMF has more low-mass clusters that are not as efficient in destroying binaries. Note that $M_{\rm scp}$ is the same value as in equation~(\ref{eq:deltat}), that is the mass of a fully populated star cluster population.

However, not only $\beta$, but also the value for $M_{\rm ecl,min}$ is set directly by the user in \BiPoS. The default value for $M_{\rm ecl,min}$ is $5 \, {\rm M}_{\odot}$, corresponding to Taurus-Auriga-like embedded star-clusters (e.g. \citealt{Joncour2018}). On the other hand, $M_{\rm ecl,max}$ is set indirectly by the user by the choice of the SFR, because $M_{\rm ecl,max}$ is calculated with equation~(\ref{eq:SFR-clustermass}) from the SFR. Note that \BiPoS$\ $only produces output tables for $M_{\rm ecl,min} < M_{\rm ecl,max}$ and leaves a failure notice on the computer screen otherwise.

If however $M_{\rm ecl,min} < M_{\rm ecl,max}$, \BiPoS$\ $evaluates $\Omega_{\rm dyn}^{\rho_{\rm ecl}}(\log_{10}(E_{\rm b}),t)$, that is the operator that determines how many binaries survive in every energy bin. For this, \BiPoS$\ $chooses $M_{\rm ecl,min}$ as the initial mass of a star cluster, and calculates $\Omega_{\rm dyn}^{\rho_{\rm ecl}}(\log_{10}(E_{\rm b}),t)$ for it. \BiPoS$\ $ then increases the mass of the star cluster in small steps of $10 \, {\rm M}_{\odot}$ and calculates $\Omega_{\rm dyn}^{\rho_{\rm ecl}}(\log_{10}(E_{\rm b}),t)$ iteratively until $M_{\rm ecl,max}$ is reached. The embedded star cluster half-mass radius, $r_h$, is taken from
\begin{equation}
\label{eq:rh-mass}
\frac{r_{\rm h}}{{\rm pc}}=0.10 \times \left( \frac{M_{\rm ecl}}{{\rm M}_{\odot}} \right)^{0.13},
\end{equation}
which is a weak mass-radius relation for embedded star clusters taken from \citet{Marks2012a}. Equation~(\ref{eq:rh-mass}) is required to calculate $\rho_{\rm ecl}$, which enters the stellar-dynamical operator. Thus, ultimately, $\rho_{\rm ecl}$ increases with increasing $M_{\rm ecl}$. The density $\rho_{\rm ecl}$ is however the only parameter relevant in determining $\Omega_{\rm dyn}^{\rho_{\rm ecl}}$ and, thus, the resulting binary fraction. The user may choose to deviate from this default behaviour by setting a constant initial value for $r_h$ for all star cluster masses (Section~\ref{sec:working}).

Equation~(\ref{eq:rh-mass}) is a theoretical result, which is needed for the dynamical population synthesis. An analogy is the stellar IMF, which is also not observable, but a mathematical workaround needed for calculations of stellar populations \citep{Kroupa2018}. The reation between radii and stellar masses at birth can be physically interpreted as the state of deepest cloud-core collapse. This corresponds to the assumption that all binaries materialised simultaneously at time zero, which is when the N-body simulation of the embedded cluster with the initial binary population begins, that is the binary population after internal binary population, or pre-main sequence eigenevoulution. Interesting is that the implied relation between densities and masses is in good agreement with the observed cloud core densities (figure~6 in \citealt{Marks2012a}). 

In order to account for the different numbers of embedded star clusters in each interval $[M_{\rm ecl},M_{\rm ecl}+10 \, {\rm M}_{\odot}]$, the resulting populations for a star cluster of a given mass are weighted with the ECMF (equation~\ref{eq:ECMF}). The numbers in each interval vary depending on the slope $\beta$ -- the larger $\beta$, the fewer embedded star clusters with increasing $M_{\rm ecl}$ in the ECMF.

The period of time, for which $\Omega_{\rm dyn}^{\rho_{\rm ecl}}(\log_{10}(E_{\rm b}),t)$ is evaluated, is fixed to a value of 3~Myr, instead of leaving the choice between 1~,3~and 5~Myr to the user, as possible in the star cluster mode in \BiPoS. The idea behind this is that most star clusters do not survive gas expulsion, that is the event that turns an embedded cluster into an open cluster, and gas expulsion has usually taken place before 5~Myr \citep{Lada2003}. Thus, after 3~Myr, most binaries are in fact part of the field of the galaxy in question, and do not interact with each other any more (see e.g. the models calculated by \citealt{Kroupa2001b}). But even star clusters that do survive gas expulsion do not loose many binaries any more, because the star clusters expand during gas expulsion, such that close interactions between binaries, which are the reason for the destruction of binaries, happen quite rarely afterwards. Also, most of the dynamical evolution of the binaries is over after 3~Myr anyway, especially for more massive embedded star clusters (see \citealt{Marks2012a} and Section~\ref{sec:SCexample}).

Note that the early application of the galactic field mode of \BiPoS$\ $led to the prediction that low-mass dwarf irregular galaxies should have field binary fractions of about 80 precent, while massive elliptical galaxies should have about 30 percent \citep{Marks2011b}. If $0.1 \, {\rm pc} \apprle r_{\rm h} \apprle \, 0.2 \, {\rm pc}$ for the half-mass radii of star clusters in the Milky Way initially, then the binary faction of the Milky Way today is about 50 percent \citep{Marks2011b}, as is observed by \citet{Duquennoy1991}.

\section{Working with \BiPoS}
\label{sec:working}

\subsection{Setup}
\label{sec:setup}

The program files of \BiPoS$\ $can be downloaded from GitHub under the web address \texttt{https://github.com/JDabringhausen/BiPoS1}. The user should take care that the folder structure is preserved when extracting \BiPoS, because the folders \texttt{Lib/} and \texttt{output/} are vital for the storage of the output data, and are not created during runtime. The program is compiled by typing \texttt{make BiPoS} into the command line of a terminal in the directory where the user has stored the programme files. \BiPoS$\ $is started afterwards by typing \texttt{./BiPoS} into the command line.

\subsection{Working with \BiPoS$\ $from the command line}
\label{sec:command-line}

After typing \texttt{./BiPoS} into the command line, the user is directed to the help menu. The syntax to be used for working with \BiPoS$\ $is explained in detail under the different items of that menu.
These items are:
\begin{itemize}
\item \texttt{./BiPoS$\ $genlib help}. With \texttt{./BiPoS$\ $genlib (...)}, the user can create a library of binary systems, which is used in later calls of \BiPoS. First-time users have to generate a library of binaries first, to which \BiPoS$\ $can refer in the following steps. 
\item \texttt{./BiPoS$\ $clust help}. With \texttt{./BiPoS$\ $clust (...)}, the user can synthesize a binary population in a single star cluster.
\item \texttt{./BiPoS$\ $field help}. With \texttt{./BiPoS$\ $field (...)}, the user can synthesize a galaxy-wide field population.
\item \texttt{./BiPoS$\ $SpT help}. Typing this into the command line shows the default mass ranges for spectral types and a way to change them.
\end{itemize}
Proficient users may skip the help menu and type in directly the syntax for the binary population they want \BiPoS$\ $to synthesize.

Some remarks are useful when working with \BiPoS:
\begin{itemize}
\item In principle, the commands \texttt{./BiPoS$\ $clust} and \texttt{./BiPoS$\ $field} suffice to put \BiPoS$\ $to work, after a standard library of binaries has been created calling \texttt{./BiPoS$\ $genlib}. It then produces the binary population of a single default star cluster, or default galactic field, respectively, where all parameters the user may specify are set to their defaults. The default for a specific value is overridden by a specification by the user.
\item \BiPoS$\ $understands only the decimal notation for numbers. Thus, for example, the user will have to use \texttt{1000} instead of \texttt{1e3}, \texttt{1e+03}, or similar notations. In the previous example, \BiPoS$\ $will break reading at the \texttt{e}; that is it interprets \texttt{1e3} or \texttt{1e+03} as \texttt{1}.
\item \BiPoS$\ $ignores parts of a command it does not understand and replaces them by the default values. It does not produce an error message when a command is wrong. Thus, the user should check the spelling of the part of the command if \BiPoS$\ $continues to use a default value.
\item The order of the parts of a command does not matter in \BiPoS. Thus, for instance, the user could type \texttt{./BiPoS$\ $clust OPD=P SpT=A libname=X} or \texttt{./BiPoS$\ $clust libname=X SpT=A OPD=P}, and \BiPoS$\ $would return the distribution of the periods of A-star binaries after $t$=3 Myr of dynamical evolution in a star cluster with $M_{\rm ecl}= 10^3 \, {\rm M}_{\odot}$ and $r_{\rm h}=0.24 \, {\rm pc}$ (that is the defaults for $t$, $M_{\rm ecl}$ and $r_{\rm h}$; the half-mass radius results from eq.~\ref{eq:rh-mass}) in both cases, while using the library of binaries \texttt{X} as a reference.
\end{itemize}

Also the ranges and the number of bins in which the binary parameters (like the periods, the semimajor axes, and so on) are divided are pre-defined by default values. However, the user can override the default for any parameter considered in \BiPoS$\ $by typing \texttt{constrain=par,x,y,z}. In this command, \texttt{par} defines the parameter to be constrained, \texttt{x} and \texttt{y} are the lower and the upper limits of the constraint, respectively, and \texttt{z} gives the number of (equal-sized) bins that the constrained zone is divided into. The parameter constrained needs not to be equal to the parameter that is outputted. Thus, for instance, the combination \texttt{OPD=a constrain=q,0.2,0.4,4} is possible. In this case, the programme outputs the distribution of semi-major axes $a$, but only for those binaries that additionally have mass-ratios ranging from $0.2$ to $0.4$. The $z=4$ bins in $q$ would only impact an outputted mass-ratio distribution, but has no effect on the here requested semi-major axis distribution. In this example an arbitrary value can be used. The user can constrain multiple parameters at the same time by using \texttt{constrain=par,x,y,z} several times in one command. The output can thus be taylored precisely to the observational constraints of the survey the results of \BiPoS$\ $are to be compared to. The user then lets \BiPoS$\ $create tables by typing \texttt{constrain=par,x,y,z} once for each parameter \texttt{par} to be constrained.

A similar scheme exists also for the spectral types of stars. By default, \BiPoS$\ $returns the binary populations of the requested parameters for stars of all spectral types in one table. If the user was instead only interested in the periods of, say, the G-stars ($0.8 \, {\rm M}_{\odot} <m \le 1 \, {\rm M}_{\odot}$) and K-stars ($0.5 \, {\rm M}_{\odot} <m \le 0.8 \, {\rm M}_{\odot}$) instead of all stars, the user would add \texttt{SpT=GK} to the command. \BiPoS$\ $then returns two tables for each requested parameter, namely one containing the distribution of just the G-stars for each requested parameter, and one with the same values for just the K-stars. The user may also choose a user-defined mass range by typing \texttt{SpT=u mmin=low mmax=up}. Here, the user would replace \texttt{low} with the lower mass limit for stars in ${\rm M}_{\odot}$, and \texttt{up} with the upper mass limit in ${\rm M}_{\odot}$ for the stars to be considered.

Note that constraining the sample to the values an observer has actually targeted may be the solution when \BiPoS$\ $finds by default unexpectedly many binaries, or unexpectedly few binaries.

For instance, by default, \BiPoS$\ $considers binaries with semi-major axes of $-5 \le  \, \log_{10}(a / {\rm AU})\le 5$. If however the observational constrains are such that, say, binaries with $\log_{10}(a / {\rm AU}) \le -1$ cannot be resolved, while the observational field is too crowded to identify binaries with $\log_{10}(a / {\rm AU}) \ge 3$ reliably, then the user may add \texttt{constrain=a,-1,3,z} to \texttt{OPD=a} in the command which sets off the calculation of $a$ with \BiPoS. \BiPoS$\ $then returns only the binaries with $-1 \le  \, \log_{10}(a / {\rm AU})\le 3$ in $z$ equal-sized bins, while it does not consider binaries with $\log_{10}(a / {\rm AU}) < -1$ or $\log_{10}(a / {\rm AU}) > 3$.
 
Likewise, the user may not be able to see (very faint) M-dwarfs below, say, $0.3 \, {\rm M}_{\odot}$ while stars more massive than, say, $2 \, {\rm M}_{\odot}$ have already evolved into remnants. However, the library the user uses has been generated for, say, stars from $0.1 \, {\rm M}_{\odot}$ to $10 \, {\rm M}_{\odot}$, meant to resemble the initial population of the cluster or galaxy field under consideration. If the user does not use any constrains, the full library from $0.1 \, {\rm M}_{\odot}$ to $10 \, {\rm M}_{\odot}$ is used for creating the resulting orbital-parameter distributions. If however the user would add \texttt{SpT=u mmin=0.3 mmax=2} to the command which sets off \BiPoS, only binaries with companion masses between $0.3 \, {\rm M}_{\odot}$ and $2 \, {\rm M}_{\odot}$ are considered in the requested semi-major axis distribution. If furthermore the user would also add \texttt{scale=N} to the command which sets off \BiPoS, where \texttt{N} is the total number of targets observed and searched for components, then the total number of binaries with primary masses between $0.3 \, {\rm M}_{\odot}$ and $2 \, {\rm M}_{\odot}$ expected in the model is returned in the third column of the output table (see end of Section~\ref{sec:impGenRemarks}).

\section{Examples}
\label{sec:examples}

Examples for the earliest usage of \BiPoS$\ $can be found in \citet{Marks2011b} for galactic fields and in \citet{Marks2012a} for star clusters. The papers also compare the results from \BiPoS$\ $with observed data. However, those papers do not mention \BiPoS$\ $explicitly. Therefore, it is introduced here, including some further examples with an emphasis on the usage of \BiPoS.

\subsection{Birth binary population and initial binary population}
\label{sec:birth-initial}

\begin{figure}
\centering
\includegraphics[scale=0.85]{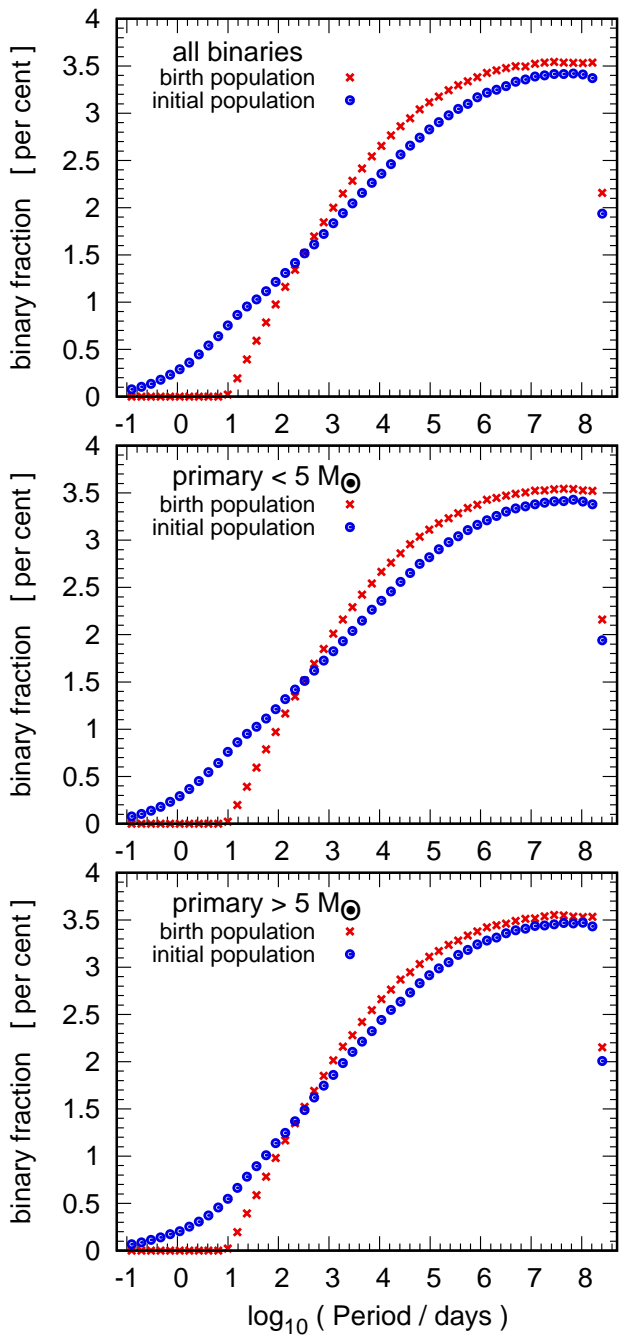}
\caption{\label{birth+init} The binary fraction per bin in percent over the periods of the binaries for the birth population (red x-symbols), and the population it evolves into through internal binary evolution, or pre-main sequence eigenevolution, namely the primordial population (blue circles). The top panel shows the birth population and the primordial population for binaries with primaries from $0.08 \, {\rm M}_{\odot}$ to $150 \, {\rm M}_{\odot}$ (that is all stars), the middle panel the birth population and the primordial population for only the binaries with primaries below $5 \, {\rm M}_{\odot}$ (that is the binaries with random pairing) and the bottom panel the birth population and the primordial population for only the binaries with primaries above $5 \, {\rm M}_{\odot}$ (that is the binaries with ordered pairing).}
\end{figure}

We want to test the effects of internal binary evolution, or pre-main sequence eigenevolution in \BiPoS. This is equivalent with checking for the difference between the birth binary population and its evolutionary descendant, the initial binary population (see also figure~1 in \citealt{Marks2011a}).

To do this, a library of binaries has to be generated. The default values for this are $10^7$ binaries with primary masses between $0.08 \, {\rm M}_{\odot}$ and $150 \, {\rm M}_{\odot}$, which are the values which we choose here. Thus, we enter \texttt{./BiPoS$\ $genlib libname=lib1.dat -eigen} for the birth binary population and \texttt{./BiPoS$\ $genlib libname=lib2.dat} for the initial binary population. Note that the initial binary population is the default in \BiPoS, because this is usually the basis for the external binary evolution, or stimulated evolution of binaries, that is the standard application of \BiPoS. Also note that \texttt{libname=libN.dat} lets \BiPoS$\ $write the contents of the library to a file stored in \texttt{Lib/libN.dat}. The default is that the library is stored in \texttt{Bin\_lib.dat}.

Also, we want to test the effect of internal binary evolution, or pre-main sequence eigenevolution, for only the binaries with primary masses of $m_1 < 5 \, {\rm M}_{\odot}$ and only the binaries with primary masses of $m_1 > 5 \, {\rm M}_{\odot}$. This is the limit where \BiPoS$\ $switches from random sampling of the binaries ($m_1 < 5 \, {\rm M}_{\odot}$) to ordered sampling ($m_1 > 5 \, {\rm M}_{\odot}$) by default. This is done by giving \BiPoS$\ $the commands \texttt{./BiPoS$\ $genlib libname=lib3.dat mmax=5.0 -eigen}, and \texttt{./BiPoS$\ $genlib libname=lib4.dat mmax=5.0}, respectively, for the library with only primary masses of $m_1 < 5 \, {\rm M}_{\odot}$, and \texttt{./BiPoS$\ $genlib libname=lib5.dat mmin=5.0 -eigen}, and \texttt{./BiPoS$\ $genlib libname=lib6.dat mmin=5.0}, respectively, for the library with only primary masses of $m_1 > 5 \, {\rm M}_{\odot}$. Note that all library sizes are still $10^7$ binaries.

The next step is to output the orbital-period distributions from the libraries before and after eigenevolution, that is to plot the birth and initial distributions. We do this by, for instance, giving \BiPoS$\ $the command \texttt{./BiPoS$\ $clust mecl=100000 OPD=P constrain=P,-1,8.5,50 +init -evolve libname=lib2.dat} for the library that contains $10^7$ internally evolved, or pre-main sequence eigenevolved, binaries with primary masses $0.08 \, {\rm M}_{\odot} \le m_1 \le 150 \, {\rm M}_{\odot}$. This command lets \BiPoS$\ $put $10^5$ binaries into 50 equal-sized bins with periods from $\log_{10}(P/{\rm days})=-1$ to $\log_{10}(P/{\rm days})=8.5$. By putting \texttt{+init} into the command, \BiPoS$\ $returns the datafiles from which it starts, that is without any external binary evolution, or stimulated evolution. By putting \texttt{-evolve} into the command, \BiPoS$\ $suppresses the datafiles with external binary evolution, or stimulated evolution.  We use the same command also for the other libraries that we mention above.

Figure~(\ref{birth+init}) shows the comparisons between the birth binary populations and the initial binary populations of all binaries (top panel), only the binaries with pimary masses $m_1  < 5 \, {\rm M}_{\odot}$ (middle panel), and only the binaries with primary masses $m_1 > 5 \, {\rm M}_{\odot}$ (bottom panel). The scaling of the binaries is such that the summing over all bins gives 100 percent binaries for both the birth binary population and the initial binary population. An alternative interpretation of this scaling is that each of the 50 bins shows the binary fraction that has periods between $P$ and $P +\Delta P$ in percent, so that the sum over all binaries is 100 percent.

Attentive observers may find that the internal binary evolution, or pre-main sequence eigenevolution, acts more strongly on the randomly sampled binaries (middle panel) than on the binaries with ordered sampling (bottom panel). However, when observing the full canonical IMF, random sampling overwhelms ordered sampling, because of the sparsity of stars with $m > 5 \, {\rm M}_{\odot}$ over all stars in the canonical IMF.

Also, the initial binary population is slightly larger than zero at the bin the most to the left. This corresponds to binaries with periods slightly larger than 0.1 days. This hints at the possibility that also binaries with periods below 0.1 days could exist, but 0.1 days is the lower limit for periods in \BiPoS. However, binaries this close would probably merge to single stars. But this concerns only a minority of binaries anyway.

Finally, after a steady rise in binary numbers with increasing periods, there is a large drop in the right-most bin. This may be surprising, because the function from which the periods stem (see equation~\ref{eq:dist-a}) only rises over the whole range where it is defined. However, the last bin in Figure~(\ref{birth+init}) stretch over the definition range of equation~(\ref{eq:dist-a}), and therefore are not fully filled with binaries. But this issue can be remediated by additionally constraining the upper limit in \BiPoS$\ $accordingly.

\subsection{Binary period distributions of star clusters in dependency of their mass and age}
\label{sec:SCexample}

\begin{figure}
\centering
\includegraphics[scale=0.89]{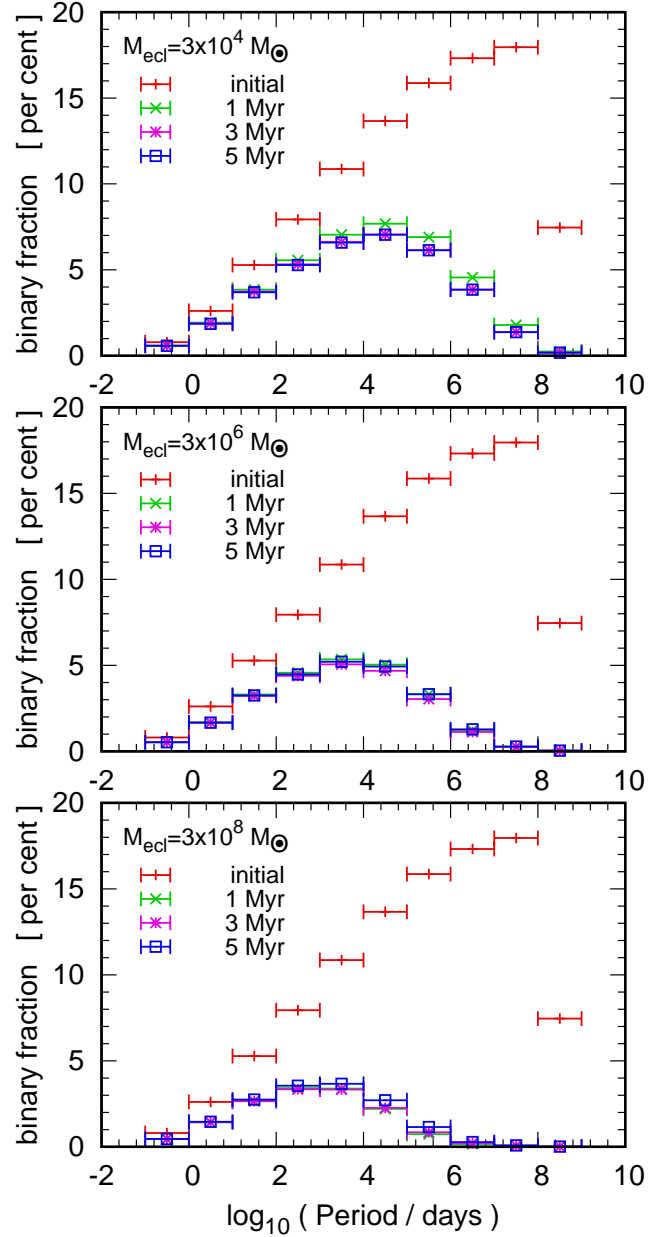}
\caption{\label{mass-age} The binary fraction per bin in per cent over the periods of the binaries for different ages. The panels show from top to bottom a star cluster with an embedded cluster mass in stars of $M_{\rm ecl}=3\times 10^4 \, {\rm M}_{\odot}$, $M_{\rm ecl}=3\times 10^6 \, {\rm M}_{\odot}$ and $M_{\rm ecl}=3\times 10^8 \, {\rm M}_{\odot}$. In each panel, the histogramms show which binary fraction survived external binary evolution, or stimulated evolution, for 0~Myr (that is 100 per cent over all bins), 1~Myr, 3~Myr and 5~Myr.}
\end{figure}

\begin{figure}
\centering
\includegraphics[scale=0.89]{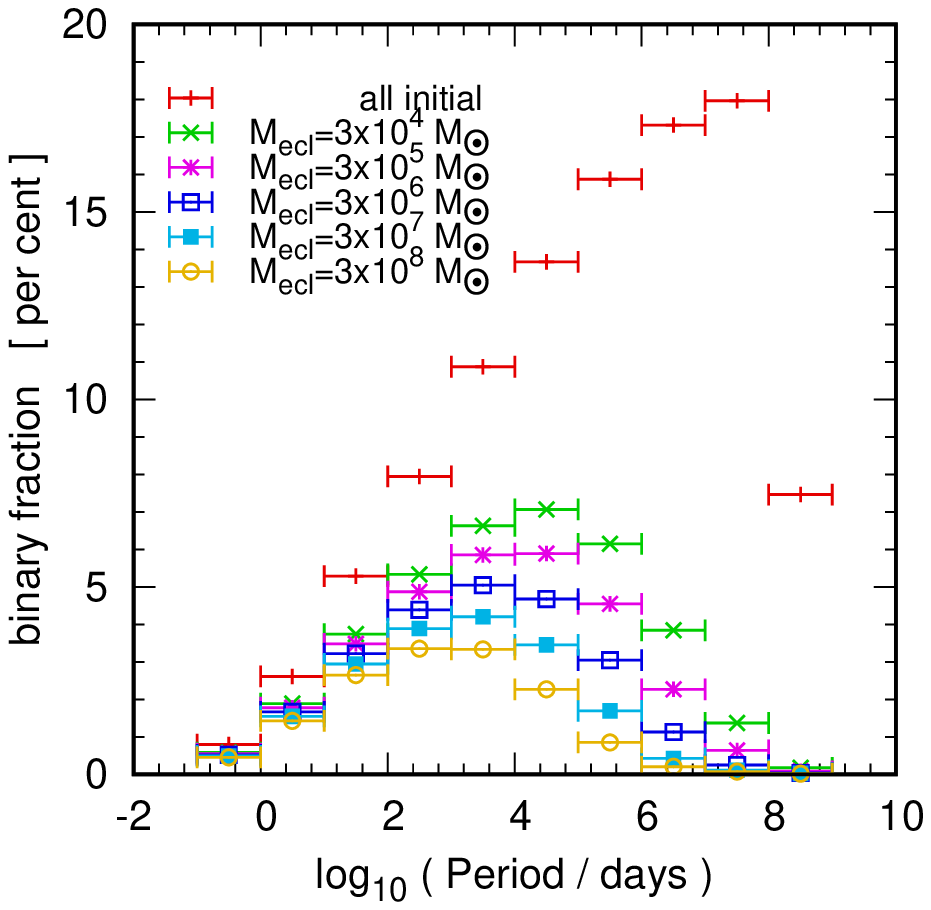}
\caption{\label{diff-mass} The binary fraction per bin in per cent over the periods of the binaries for different masses. The top curve is the initial binary distribution, which is for all star clusters the same due to the universality of not only the IMF \citep{Kroupa2001a} (except under extreme conditions, see e.g. \citealt{Dabringhausen2012,Marks2012b,Kroupa2013}), but also of the binary distribution function \citep{Marks2012a} (see Section~\ref{sec:discussion} for a discussion). The curves below show from top to bottom how the binary population has evolved after 3~Myr, if it is in an embedded cluster mass of$ M_{\rm ecl}=3\times 10^4 \, {\rm M}_{\odot}$, $M_{\rm ecl}=3\times 10^5 \, {\rm M}_{\odot}$, $M_{\rm ecl}=3\times 10^6 \, {\rm M}_{\odot}$, $M_{\rm ecl}=3\times 10^7 \, {\rm M}_{\odot}$ and $M_{\rm ecl}=3\times 10^8 \, {\rm M}_{\odot}$ from top to bottom.}
\end{figure}

We want to check the dynamical evolution of the periods of the binary population in star clusters with initial total embedded stellar masses, $M_{\rm ecl}$, from $3\times 10^4 \, {\rm M}_{\odot}$ to $3\times 10^8 \, {\rm M}_{\odot}$. Such star clusters are thought to be the precursors of globular clusters. We assume no peculiarities about the lower mass limit of the stars in star clusters, $m_{\rm min}$, so that it is at $0.08 \, {\rm M}_{\odot}$ for all of them (e.g. \citealt{Thies2007}). The upper stellar mass limit of star clusters, $m_{\rm max}$, saturates at about $150 \, {\rm M}_{\odot}$ for star clusters with total masses $M_{\rm ecl} \apprge 10^4 \, {\rm M}_{\odot}$ (e.g. \citealt{Weidner2010}). Therefore, we generate the default library in \BiPoS$\ $(see Section~\ref{sec:birth-initial}) and call it \texttt{GCs.dat}.

Note that the library size ($10^7$ binaries) is smaller than required for the largest cluster masses here. With the full canonical IMF, like it is chosen here, the star clusters with $M_{\rm ecl}=10^8 \, {\rm M}_{\odot}$ must also have more stars than there are stars in the library of binaries. However, this is no issue in \BiPoS, because \BiPoS$\ $calulates the ratio of surviving binaries over the initial binaries in every binary bin, and then selects the binaries from the library until that ratio is reached.

Furthermore, we assume the average half-mass radius for each star cluster is given by equation~(\ref{eq:rh-mass}). This is equation~7 in \citet{Marks2012a} and the default assumed by \BiPoS. Finally, we calculate the surviving binary fractions for all three ages that \BiPoS$\ $offers, that is 1,~3~and 5~Myr. Thus, for instance, \BiPoS$\ $calculates the periods of the binary fraction for a cluster with $M_{\rm ecl}=3\times 10^4 \, {\rm M}_{\odot}$, a radius following from equation~(\ref{eq:rh-mass}) and a binary evolution time of 3 Myr after typing \texttt{./BiPoS$\ $clust OPD=P mecl=30000 libname=GCs.dat} into the command line. We request a distribution of periods by typing \texttt{OPD=P}, for a star cluster with $M_{\rm ecl}=3\times 10^4 \, {\rm M}_{\odot}$ by typing \texttt{mecl=30000}, and add \texttt{libname=GCs.dat} for the name we have given the library of binaries we have created before. For the other values, we are content with the default values. Figures~(\ref{mass-age}) and~(\ref{diff-mass}) compare the period distribution functions of the binary stars for the calculated star clusters with each other.

Figure~(\ref{mass-age}) shows a star cluster with an embedded cluster mass of $M_{\rm ecl}=3\times 10^4 \, {\rm M}_{\odot}$ (upper panel), $M_{\rm ecl}=3\times 10^6 \, {\rm M}_{\odot}$ (middle panel) and $M_{\rm ecl}=3\times 10^8 \, {\rm M}_{\odot}$ (lower panel) for 0,~1,~3~and 5~Myr of external binary evolution, or stimulated evolution, of the periods of the binaries. The y-axis shows the percentage of the binaries per period bin. Thus, the sum over all bins together shows the initial fraction of binary stars, which is 100 percent in \BiPoS. It is shown here as the uppermost red lines. Also, the red lines are identical for all three panels, which is a consequence of the universality of the binary distribution function \citep{Kroupa2011IAU,Marks2012a}, which is adopted in \BiPoS. The sum over all bins for 1,~3, and~5 Myr of external binary evolution, or stimulated evolution, show the percentage of binaries which survive this time of dynamical evolution. This is 37 percent of the binaries for a cluster with $M_{\rm ecl}=3\times 10^4 \, {\rm M}_{\odot}$, 30 percent of the binaries for a cluster with $M_{\rm ecl}=3\times 10^5 \, {\rm M}_{\odot}$, 24 percent of the binaries for a cluster with $M_{\rm ecl}=3\times 10^6 \, {\rm M}_{\odot}$, 19 percent of the binaries for a cluster with $M_{\rm ecl}=3\times 10^7 \, {\rm M}_{\odot}$ and 15 percent of the binaries for a cluster with $M_{\rm ecl}=3\times 10^8 \, {\rm M}_{\odot}$ for an age of $t=3$~Myr, but in fact almost independent of which one of the three ages was chosen. This indicates that almost all external binary evolution, or stimulated evolution, for the shown star clusters happens before the smallest choosable dynamical evolution time, which is 1~Myr. Figure~(\ref{diff-mass}) thus only shows the period distributions of clusters with different $M_{\rm ecl}$ after 3~Myr of external binary evolution, or stimulated evolution. That the binary distributions are independent of whether 1~Myr, 3~Myr or 5~Myr was chosen as the age of the star cluster is however not true for small star clusters, where the external binary evolution takes much longer until it is stopped by gas expulsion, as can be seen in \citet{Marks2012a}.

While \BiPoS$\ $asks for $M_{\rm ecl}$ and $r_{\rm h}$ on the input, it uses the initial density, $\rho_{\rm ecl}$, to calculate its output. The combinations of $M_{\rm ecl}$ and $r_{\rm h}$ in this example imply a rising $\rho_{\rm ecl}$ with raising $M_{\rm ecl}$. A higher density in turn implies a more efficient binary destruction, until only the most tightly bound binaries survive. Thus, with all the other parameters the same, that is, the same library of binaries was used, and \BiPoS$\ $always assumes a binary fraction of 100 percent for the initial population, the density of a star cluster decides which percentage of binaries survive dynamical evolution for a time $t$. This happens in a time span of less than a Myr in the case of massive and dense star clusters, like here. In consequence, N-body simulations of young GCs with a very low binary content like the ones in \citet{Wang2016} are justified, as is demonstrated here with \BiPoS.

\subsection{Close binaries in the Orion Nebula Cluster}
\label{sec:ONC}

\begin{figure}
\centering
\includegraphics[scale=1.0]{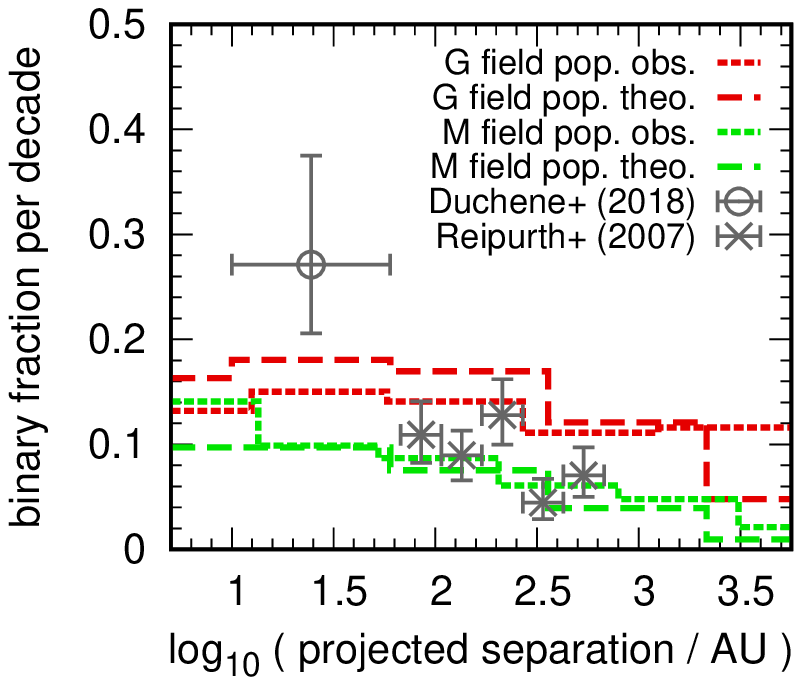}
\caption{\label{fig:duchene2018-0} The observed binary fractions for the ONC normalized by the bin-width over apparent separation. The data from \citet{Reipurth2007} are shown as grey x-symbols and the data from \citet{Duchene2018} is shown as grey circle. The red dotted lines are the observations for M-type field binaries by \citet{WardDuong2015} and the green dotted lines are the observations for G-type field binaries by \citet{Raghavan2010}. The dashed lines are the field populations according to \BiPoS$\ $for M-type primaries in green, and for G-type primaries in red.}
\end{figure}

\begin{figure}
\centering
\includegraphics[scale=1.0]{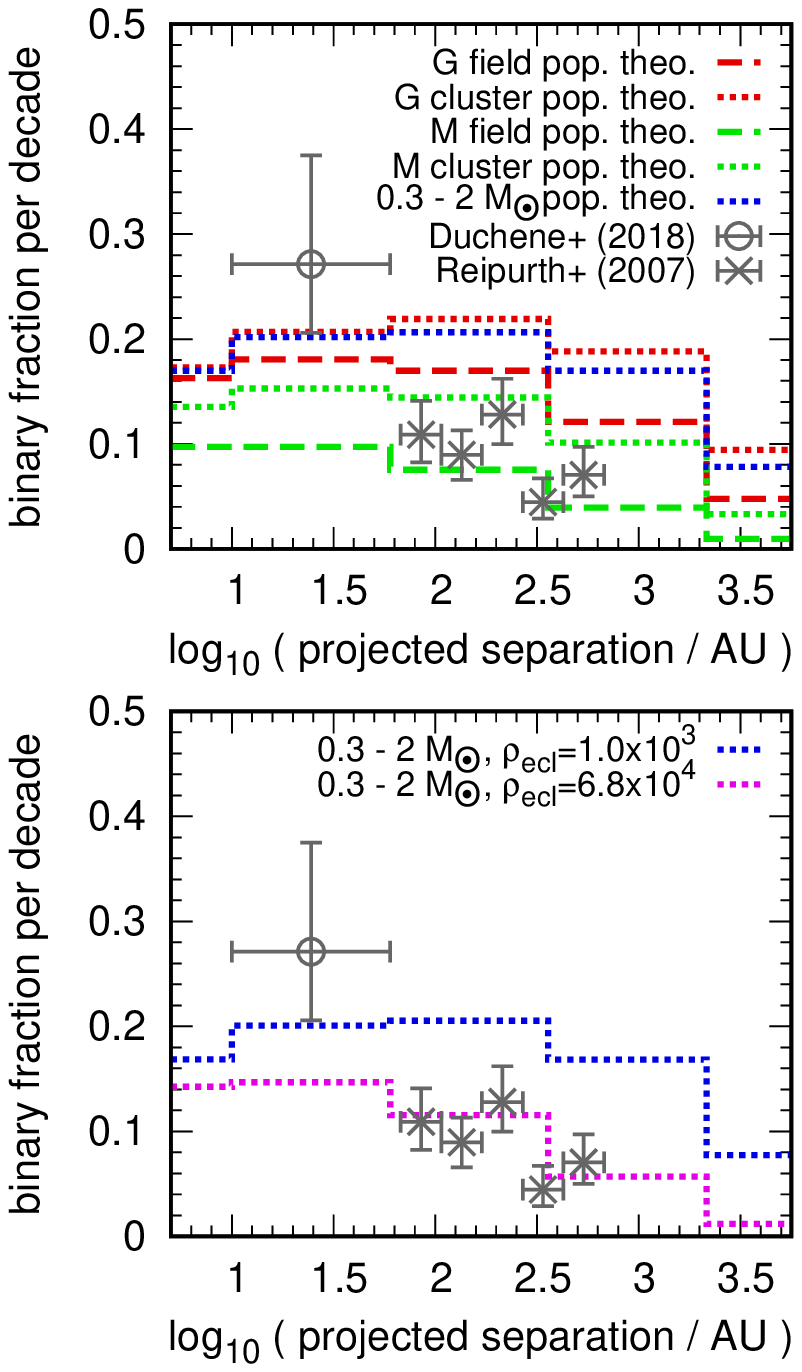}
\caption{\label{fig:duchene2018-1} The observed binary fractions for the ONC normalized by the bin-width over apparent separation. The data from \citet{Reipurth2007} are shown as grey x-symbols and the data from \citet{Duchene2018} are shown as grey circle. They are compared to different binary fractions calculated with \BiPoS. In the upper panel, the binary fractions according to \BiPoS$\ $are the field population of the M-type primaries indicated as green dashed line, and the field-population of G-type primaries indicated as red dashed line (like in Fig.~\ref{fig:duchene2018-0}). The dotted lines are the binary fractions for the ONC with present-day parameters according to \citet{Hillenbrand1998}; again for M-type primaries in green and for G-type primaries in red. The blue dotted line is also for the ONC with present-day parameters, but for primary masses of $0.3 \, {\rm M}_{\odot} le m_1 \le 2 \, {\rm M}_{\odot}$. It is also shown in the lower panel, but here indicated with the initial density $\rho_{\rm ecl}=1050 \, {\rm M}_{\odot} \, {\rm pc}^{-3}$, as these are the present-day parameters for the cluster mass and half-mass radius from \citet{Hillenbrand1998} converted into density. Density is the parameter that is actually relevant for \BiPoS. This is compared to a initial density of $\rho_{\rm ecl}=68000 \, {\rm M}_{\odot} \, {\rm pc}^{-3}$, which is indicated by a violet dotted curve. $\rho_{\rm ecl}=68000 \, {\rm M}_{\odot} \, {\rm pc}^{-3}$ is the most likely initial density for the ONC according to \citet{Marks2012a}.}
\end{figure}

\citet{Duchene2018} state that the Orion Nebula Cluster (ONC) has a companion star fraction in close binaries that is about twice as high as in comparable field stars. While this is true, it cannot be concluded that the initial binary function is not universal, as we demonstrate in this section.

Note that in figures~(\ref{fig:duchene2018-0}) and~(\ref{fig:duchene2018-1}) in this Section, the binary fractions per decade are shown on the y-axis, and not just the plain binary fractions like in Sections~(\ref{sec:birth-initial}),~(\ref{sec:SCexample}) and~(\ref{sec:small}). Thus, summing over all bins does not give the total fraction of binaries over centre-of-mass systems here. The advantage is however that also companion star fractions that were obtained for different bin-widths can be directly compared to each other. Also, these are the numbers that appear in the second columns of the output from \BiPoS$\ $for this reason.

Figure~(\ref{fig:duchene2018-0}) shows the datapoint for the companion star fraction between 10 and 60 AU in the ONC from \citet{Duchene2018} (grey circle), and the datapoints from \citet{Reipurth2007} (grey x-symbols). The datapoints from \citet{Reipurth2007} are also for companion star fractions in the ONC, but they have wider separations. Also the observed binary fractions for M-type field stars \citep{WardDuong2015} and G-type field stars \citep{Raghavan2010} are shown in this figure. They are compared in this figure with the according results from \BiPoS, which were obtained by entering the command \texttt{./BiPoS$\ $field SpT=M OPD=s constrain=s,0.222,4.112,5} into \BiPoS. Here, \BiPoS$\ $generates apparent separations, $s$, of a field population of binaries with M-type stars as primaries. The part \texttt{constrain=s,0.222,4.112,5} of the command lets \BiPoS$\ $divide the range in $s$ from $\log(s/{AU})=0.222$ to $\log(s/{AU})=4.112$ into five equal-sized bins. This was inspired from the filled data point in figure~(6) in \citet{Duchene2018}, or the grey circle in Fig.~(\ref{fig:duchene2018-0}) here. It is matched perfectly by the second bin here, while the other bins cover the full range of data in $s$. The same command was also given for G-stars, except that \texttt{SpT=M} was replaced by \texttt{SpT=G} for them.

\BiPoS$\ $ thus only picks stars with masses $0.08 \, {\rm M}_{\odot}< m_1 \le 0.45 \, {\rm M}_{\odot}$ for the M-stars from the user-generated library, and with masses $0.8 \, {\rm M}_{\odot}< m_1 \le 1 \, {\rm M}_{\odot}$ for the G-stars. Observationally, it is doubtful whether the stars are complete in \citet{Raghavan2010}, and in \citet{WardDuong2015}, respectively. On the other hand, the sample in \citet{Raghavan2010} contains also some low-mass K-stars, and the sample in \citet{WardDuong2015} contains also some F-stars according to the definitions used in the papers and in \BiPoS. Thus, it cannot be expected that the comparisons between the observations and \BiPoS$\ $ would match precisely, but they can at least give a hint.

Nevertheless, the data from \BiPoS$\ $indeed match the observed data from \citet{WardDuong2015}, and \citet{Raghavan2010}, respectively, very well. Also the match of the observed data from \citet{Reipurth2007} with the M-type field binaries is good, while they are a bit too low for G-type field stars. The match with M-type stars is remarkable, because the data from \citet{Reipurth2007} is actually taken in the ONC, and not the field. The data point from \citet{Duchene2018} on the other hand lies much higher than both the observed and calculated values for field stars.

We now enter the information that the ONC is a star cluster with specific properties into \BiPoS. According to \citet{Hillenbrand1998}, the ONC has an $r_{\rm h}$ of 0.8 pc \textit{today}, a total mass of about 4500 ${\rm M}_{\odot}$ and is $\apprle$1 Myr old. This mass implies a density within $r_{\rm h}$ of 4400 ${\rm M}_{\odot} \, {\rm pc}^{-3}$. Thus, we enter \texttt{./BiPoS$\ $clust SpT=M OPD=s contrain=s,0.222,4.112,5 mecl=4500 rh=0.8 t=1} for M-stars into \BiPoS. Note that \texttt{field} of the above command was replaced by \texttt{clust}, in order to tell \BiPoS$\ $to synthesize a star cluster population. $M_{\rm ecl}$, $r_{\rm h}$ and age of the star cluster are chosen as close as possible to the data in \citet{Hillenbrand1998}. Also here, the command was entered again for G-stars.

Moreover, \citet{Duchene2018} have observed stars from $0.3 \, {\rm M}_{\odot}$ to $2 \, {\rm M}_{\odot}$, and not just M-type stars ($0.08 \, {\rm M}_{\odot}$ to $0.5 \, {\rm M}_{\odot}$) or just G-type stars ($0.8 \, {\rm M}_{\odot}$ to $1 \, {\rm M}_{\odot}$). This can be accomodated for in \BiPoS$\ $by replacing \texttt{SpT=M} with \texttt{SpT=u mmin=0.3 mmax=2}, which lets \BiPoS$\ $look for primaries with masses from $0.3 \, {\rm M}_{\odot}$ to $2 \, {\rm M}_{\odot}$ instead of M-stars. Note that also in this case, the data for observed stars is hardly complete, in contrast to the stars with $0.3 \, {\rm M}_{\odot}< m_1 \le 2 \, {\rm M}_{\odot}$ in \BiPoS. However suprisingly, the resulting histogram very much resembles the histogram for just the G-stars, even though the range of stars encompasses also K-stars and bright M-stars at low stellar masses, and F-stars and A-stars at high stellar masses.

Finally, the ONC may be in the process of gas expulsion, and was therefore very likely denser when it formed \citep{Kroupa2001b}. According to figure~(3) in \citet{Marks2012a}, the most likely initial density of the ONC was 68000 ${\rm M}_{\odot} \, {\rm pc}^{-3}$, and not 1050 ${\rm M}_{\odot} \, {\rm pc}^{-3}$, as it is today. Thus, we replace \texttt{mecl=4500 rh=0.8} with \texttt{rho=68000} in the above command. 

The results with \BiPoS$\ $for the most likely parameters of the ONC are seen as the violet curve in the lower panel of Fig.~(\ref{fig:duchene2018-1}). It is still more than 1$\sigma$ below the observational value, but definitively much less than 2.7$\sigma$, as for the M-type field stars observed by \citet{WardDuong2015}. This is not a problem by itself, and could happen by pure chance, even if the model in \BiPoS$\ $were correct. However, \citet{Marks2014} noted that also other star clusters show an excess of close, very low-mass binaries, if the observations are compared to \BiPoS. This may be understood with brown dwarfs interspersed to the stars at low stellar masses. The brown dwarfs would essentially form as very massive planets in close vicinity of their primary star, while actual formation of low-mass stars would take place farther away. This problem does not exist with wider binaries, because then also for very low-mass binaries, the wider `star-like' formation mode predominates over the closer `brown-dwarf-like' formation mode. \BiPoS$\ $assumes however that all stars, also those with extremely low masses, form in the `star-like' mode. The ONC was not observed at separations comparable with closer star clusters until \citet{Duchene2018}. However, now that it is, the binary excess appears also here. We return to the issue of the `brown-dwarf-like' stellar population in Section~(\ref{sec:BD}). On the other hand, the data for the observational binary fractions at wider separations from \citet{Reipurth2007} fit the results from \BiPoS$\ $taylored to the observations by \citet{Duchene2018} very well.

The ONC has also been found to be more complex with the discovery that it consists of three subpopulations of different ages \citep{Beccari2017}. This is supported by \citet{Jerabkova2019} with data from Gaia. \citet{Jerabkova2019} also discovered evidence for wide binaries in the ONC, which most likely belong to the youngest, just-born population. Also, much more stellar dynamical modelling is needed, which takes the removal of photo-ionising stars through ejection into account \citep{Kroupa2018b, Wang2019}.

\subsection{Two small star-forming regions in Orion}
\label{sec:small}

\begin{figure}
\centering
\includegraphics[scale=0.85]{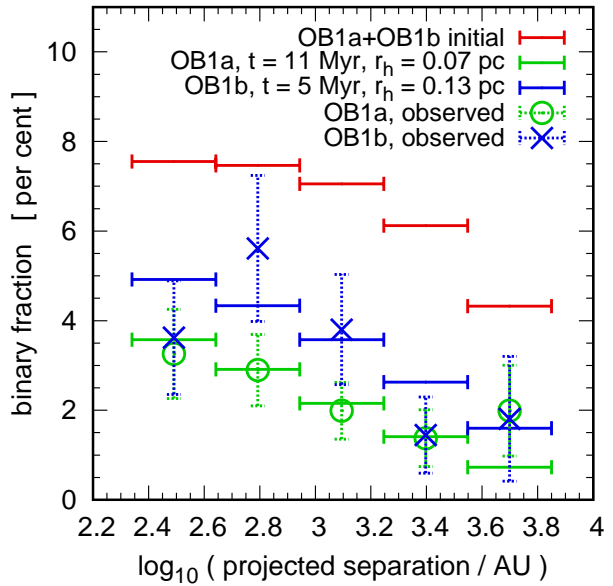}
\caption{\label{fig:OB1a+OB1b} The observed binary fractions per bin of the stellar associations OB1a (green circles) and OB1b (blue exes) according to \citet{Tokovinin2020} over apparent separation. Shown as horizontal lines of the same colours are correspondent field populations according to \BiPoS. The top (red) lines are the initial binary distribution, which is for both stellar associations almost the same. Seen in the observed range are about 32.6 percent of all initial binaries (=100 percent).}
\end{figure}

OB1a and OB1b are stellar associations in Orion (see \citealt{Blaauw1964} and \citealt{Warren1977} for details on them). \citet{Tokovinin2020} consider in their paper 658 centre-of-mass (CMS) systems in OB1a and 363 CMS systems in OB1b. These CMS systems can either be single pre-main-sequence (PMS) stars or they can have a companion star. The primary stars (whether they turn out to be singles or binaries) are in the mass range $0.3 \, {\rm M}_{\odot} <m_1 <0.9 \, {\rm M}_{\odot}$. We assume for simplicity that the primaries are complete in this mass range, although the truth is probably more complicated than that: \citet{Tokovinin2020} state that PMS stars with $0.4 \, {\rm M}_{\odot} <m_1 <0.8 \, {\rm M}_{\odot}$ dominate in their figure~(6), even though the figure shows PMS stars with $0.3 \, {\rm M}_{\odot} <m_1 <0.9 \, {\rm M}_{\odot}$, and they furthermore say that the mass estimates of PMS stars are highly uncertain anyway.

\citet{Tokovinin2020} detected companions to 74 PMS stars in OB1a and to 61 PMS stars in OB1b. They are thus binaries\footnote{In fact, \citet{Tokovinin2020} do not report 74+61=135 binaries, as is assumed here, but 127 binaries and 4 triples. The 4 triples can however formally be converted into 8 binaries, consisting of 4 inner binaries with the comparatively far away third star being neglegted, and 4 outer binaries with the inner two stars being treated as one star. This again leads to 135 binaries.}. \citet{Tokovinin2020} have also set the detection limit for a binary to $\Delta J <3$ mag, where $\Delta J$ is the magnitude difference of the two paired stars in the $J$-band of the 2MASS-survey. If the mass ratio of the stars, $q$, is calculated from their luminosities as $q \approx 10^{-0.3 \Delta J}$, like in \citet{Tokovinin2020}, $0.13 < q <1$ for $3 > \Delta J > 0$. This means that almost the complete range of values between 0 and 1 for $q$ is covered, and if the distribution of the values for $q$ follow from a random selection of the companion stars from the IMF in this mass range, almost every binary has been detected. Random pairing of stars from the IMF in that mass range is indeed proposed in \citet{Kroupa1993}, \citet{Leinert1993} and \citet{Kroupa1995a}, and is also implemented as default in \BiPoS. Thus, we assume that the 658 CMS systems in OB1a are 658+74=732 PMS stars and the 363 CMS systems in OB1b are 363+61=424 PMS stars in the mass range $0.3 \, {\rm M}_{\odot} <m <0.9 \, {\rm M}_{\odot}$ (now primary and secondary stars together).

A normalisation value $k$ for the IMF in OB1a and OB1b, respectively, over the considered mass range can be calculated over
\begin{equation}
N=k \int^{0.9}_{0.3} \xi(m) \, dm,
\end{equation}
where $N$ is the number of stars in OB1a and OB1b with masses $0.3 \, {\rm M}_{\odot} <m <0.9 \, {\rm M}_{\odot}$, respectively. Assuming that the PMS stars in OB1a and OB1b follow the canonical IMF, this leads to $k=617.4$ for OB1a and to $k=357.6$ for OB1b. If it is furthermore assumed that OB1a and OB1b formed stars over the full range of stellar masses from $0.08 \, {\rm M}_{\odot}$ to $150 \, {\rm M}_{\odot}$, then OB1a formed 2441 stars and OB1b formed 1414 stars. Given that the average mass of a star is $0.59 \, {\rm M}_{\odot}$ for the canonial IMF for stellar masses from $0.08 \, {\rm M}_{\odot}$ to $150 \, {\rm M}_{\odot}$, the total mass of OB1a would be roughly $M_{\rm ecl}=1400 \, {\rm M}_{\odot}$ and the total mass of OB1b roughly $M_{\rm ecl}=800 \, {\rm M}_{\odot}$.

The above numbers for $N$ and $M$ of OB1a and OB1b are only rough approximations, because OB1a and OB1b are stellar associations. This means that even before their dissolution, they were complexes of multiple small, young star clusters, and not single star clusters. For this reason, they are rather created with the field mode than with the star cluster mode in \BiPoS. We will do this in the following, but we still use the calculated values for $M$ and $N$, where needed.

According to \citet{Briceno2019}, OB1a has an age of $\approx 11$ Myr and OB1b an age of $\approx 5$ Myr, which we take also as durations for star formation in them. Thus, OB1a formed its total stellar mass of 1400 ${\rm M}_{\odot}$ with an average star formation rate (SFR) of $1.27\times 10^{-4} \, {\rm M}_{\odot} \, {\rm yr}^{-1}$, and OB1b its total stellar mass of 800 ${\rm M}_{\odot}$ with an average SFR of $1.6\times 10^{-4} \, {\rm M}_{\odot} \, {\rm yr}^{-1}$.

This can be translated into maximum stellar masses that were formed with those SFRs with the theory of integrated galactic stellar initial mass functions (IGIMFs) \citep{Kroupa2003}. According to equation~(14) in \citet{Dabringhausen2019}, the result is $m_{\rm max}=9.9 \, {\rm M}_{\odot}$ for the most massive star in OB1a and $m_{\rm max}=10.6 \, {\rm M}_{\odot}$ for the most massive star in OB1b. Thus, in retrospective, the choice of $M_{\rm scp}=1400 \, {\rm M}_{\odot}$ for OB1a and $M_{\rm scp}=800 \, {\rm M}_{\odot}$ for OB1b proves to be appropriate, because only about 0.5 per cent of the stars which formed with the full canonical IMF are more massive than $10 \, {\rm M}_{\odot}$, and they account for about 20 per cent of the total mass of the full canonical IMF.

We create binary libraries with \BiPoS$\ $by typing  \texttt{./BiPoS$\ $genlib libname=OB1a.dat mmax=9.9} for OB1a and \texttt{./BiPoS$\ $genlib libname=OB1b.dat mmax=10.6} for OB1b into the command line. For the lower limits of the libraries, we use the default values of $0.08 \, {\rm M}_{\odot}$ by not specifying any other value, and \texttt{OB1a.dat} and \texttt{OB1b.dat} are the chosen library names (see also Section~\ref{sec:birth-initial} for the creation of binary libraries).

The quantities that \citet{Tokovinin2020} searched for are the apparent projected separations of their binaries. They find values between 0.6 and 19.2 arcsec. The mean distances of OB1a and OB1b are 363 pc and 388 pc respectively, so that not much information is lost if both associations are assumed to be 370 pc away \citep{Tokovinin2020}. Hence, this distance is adopted here as well, and the measured apparent projected separations of the binaries turn into real projected distances between 222 AU and 7104 AU. \citet{Tokovinin2020} divide this range into five equal-sized bins in logarithmic distance.

Thus, as a first step, \BiPoS$\ $can predict the binary population of OB1a with the command \texttt{./BiPoS$\ $field OPD=s constrain=s,2.34,3.85,5 constrain=q,0.1,1,z SpT=u mmin=0.3 mmax=0.9 sfr=0.000127 libname=OB1a.dat}. The command-line argument \texttt{OPD=s} tells \BiPoS$\ $to put the projected separations in the output file. The argument \texttt{constrain=s,2.34,3.85,5} tells \BiPoS$\ $not to use the default values for the binning of the projected separations $s$, but to make 5 equal-sized bins for them from $\log_{10}(s/{\rm AU})=2.34$ to $\log_{10}(s/{\rm AU})=3.85$. The argument \texttt{constrain=q,0.1,1,z} limits the mass ratios from $0<q\le 1$ to $0.1\le q \le 1$. Note that the letter \texttt{z} is a dummy, because there are no output files from \BiPoS$\ $for $q$. Internally, \BiPoS$\ $interprets this letter as that the values for $q$ are not to be sorted into bins\footnote{Actually, to leave out the argument \texttt{constrain=q,0.1,1,z} would lead to almost the same results, because $q$ covers almost the complete range of values, as argued earlier.}.  The argument \texttt{SpT=u} tells \BiPoS$\ $to use a user-defined interval for stellar masses. This interval is defined with the next two parameters, \texttt{mmin=0.3} and \texttt{mmax=0.9}, which give the lower mass limit and the upper mass limit in Solar units. The argument \texttt{sfr=0.000127} defines the SFR, which according to equation~(6) in \citet{Weidner2004b} is equivalent to setting the upper mass limit for star clusters. Finally, the argument \texttt{libname=OB1a.dat} tells \BiPoS$\ $to use the library \texttt{OB1a.dat} in the folder \texttt{Lib/}, where the library of that name is stored. For the binary population of OB1b, the same syntax is used, except that \texttt{sfr=0.000127} is replaced by \texttt{sfr=0.00016} and \texttt{libname=OB1a.dat} is used instead of by \texttt{libname=OB1b.dat}.

The results are in both cases a total binary fraction of about 20 percent with \BiPoS, while \citet{Tokovinin2020} found total binary fraction of $11.2 \pm 1.3$ percent in OB1a and $16.8 \pm 2.2$. percent in OB1b. Thus, the results from \citet{Tokovinin2020} do not only show a significant difference, but are also notably lower.

Now, let us suppose that the star formation time scale was shorter than originally presumed, so that the SFR was higher, and conseqently more massive star clusters could form. However, even if the SFR was as high that $1400 \, {\rm M}_{\odot}$, and $800 \, {\rm M}_{\odot}$ in stars, respectively, could form in just 1 Myr, the total binary fractions become 17.0 percent for OB1a and 17.5 percent for OB1b.

Let us then suppose that not all stars which formed in OB1a and OB1b are seen, so that the SFR was higher still. For the default version of selecting the $r_{\rm h}$ of the star clusters in \BiPoS, namely with equation~(7) in \citet{Marks2012a}, we arrive at the observed binary fractions for an SFR of $2\times 10^{-3} \, {\rm M}_{\odot} \, {\rm yr}^{-1}$ in OB1b and at an SFR of $4 \, {\rm M}_{\odot} \, {\rm yr}^{-1}$ in OB1a. In the case of OB1a, this would be more than the actual SFR of the whole Milky Way according to \citet{Prantzos1995}! Thus, we discard this possibility.

Also changing the slope, $\beta$, of the ECMF is as ineffective as raising the SFR for solving the problem. This is no surprise, because for effectively changing the ratio between high-mass clusters and low-mass clusters, high-mass clusters have to form in the first place. High-mass star clusters form however only at a high SFR (see equation~6 in \citealt{Weidner2004b}). Thus, we go with the standard value in \BiPoS, which is $\beta=2$.

This leaves changing $r_{\rm h}$ of the star clusters as the only viable option for effectivly changing the binary fraction of OB1a and OB1b. This is not to say that all star clusters in OB1a formed with this $r_{\rm h}$ and all star clusters in OB1b formed with another, smaller $r_{\rm h}$. What is said, though, is that the star formation in the said stellar associations was equivalent to all stars forming in star clusters of that radius, while the star clusters can still have their individual $r_{\rm h}$.

This notion is not in tension with equation~(7) in \citet{Marks2012a}, or equation~(\ref{eq:rh-mass}) here, the default choice for $r_{\rm h}$ with $M_{\rm ecl}$ in \BiPoS. But it does indicate that the said equation has not only a most likely value, but also a large variance. For instance, a star cluster with $M_{\rm ecl}=1000 \,  {\rm M}_{\odot}$ has a most likely value of $r_{\rm h}=0.25 \, {\rm pc}$ according to equation~(\ref{eq:rh-mass}), but also values between $r_{\rm h}=0.11 \, {\rm pc}$ and $r_{\rm h}=0.55 \, {\rm pc}$ are possible within 1$\sigma$ of that equation. This implies significant changes for the surviving binary fractions, and shows that leaving the user the possibility to change the average $r_{\rm h}$ is an advantage. This is especially so when dealing with rather small populations, like in this section.

The results of fits-by-eye for a varying $r_{\rm h}$ can be seen in Fig.~(\ref{fig:OB1a+OB1b}). In this figure, \texttt{rh=0.07} was added to the command for \texttt{OB1a.dat}, and \texttt{rh=0.13} was added to the command for \texttt{OB1b.dat}, to fix $r_{\rm h}$ to 0.07 pc, and 0.13 pc, respectively. We compare this with the most likely $r_{\rm h}$ of a star cluster with $M_{\rm ecl}=100 \,  {\rm M}_{\odot}$, that is about the mass of the most massive star cluster with a SFR of $\approx 1.5\times 10^{-4} \, {\rm M}_{\odot} \, {\rm yr}^{-1}$. It would be about 0.18 pc according to equation~(\ref{eq:rh-mass}). On the other hand, the most likely value for $r_{\rm h}$ according to equation~(\ref{eq:rh-mass}) for a star cluster with $M_{\rm ecl}=5 \,  {\rm M}_{\odot}$ is 0.12 pc. Thus, the actual average value for $r_{\rm h}$ is still roughly consistent with the most likely value for $r_{\rm h}$ according to equation~(\ref{eq:rh-mass}) in the case of OB1a, but definitely below the most likely value for $r_{\rm h}$ in the case of OB1b.

The red lines in Fig.~(\ref{fig:OB1a+OB1b}) indicate the initial binary distribution, that is before external evolution, or stimulated evolution, of the binaries. It is almost identical for OB1a and OB1b, as the SFRs in both associations are almost the same. In general however, while both binary distributions would have a binary fraction of 100 per cent in total, their distributions would be different because of the different $m_{\rm max}$ of the stars. At the moment, there is however no reason to assume that the initial binary populations of OB1a and OB1b were different in the sense that, for example, OB1b formed with a binary fraction of 100 percent and OB1a did not. The difference they show nowadays can also be explained by them forming both with 100 percent binaries and having credible variations of their parameters at their births. These parameters are the radii and masses of the embedded clusters which contributed to the associations; see for example \citet{Megeath2016} for the type of embedded clusters which are creating the currently forming new OB association in Orion.

\section{Discussion}
\label{sec:discussion}

\subsection{Are binaries created in \BiPoS?}
\label{sec:create}

It is stated in Section~(\ref{sec:impGenRemarks}) that \BiPoS$\ $can only destroy binaries, but yet there are more binaries after 5 Myr than after 3 Myr of evolving the same star clusters with \BiPoS$\ $in some cases in Section~(\ref{sec:SCexample}). This difference is only about 1 per cent or less, when 100 per cent stands for the initial number of binaries, but undeniable.

This seeming contradiction is resolved in Section~(\ref{sec:omega}), where it is stated that the user may choose between 3 options for the considered time, and then \BiPoS$\ $looks up the parameters accordingly. The three sets of parameters $a(t)$ to $g(t)$ (see equation~\ref{eq:operator2} to~\ref{eq:operator4} for those parameters) are howewer fits to direct $N$-body simulations. In general, $N$-body simulations can also form binaries from single stars (see e.g. \citealt{Hut1992,Kroupa1995a}). This is however unlikely to be the case here, because the $N$-body simulations on which \BiPoS$\ $is based started at a binary fraction of 100 percent. Thus, binaries can only be destroyed in the beginning. When the first single stars appear in the $N$-body simulations, binaries can also be created, but they continue to be overwhelmed by the ongoing destruction of binaries. Later on, an equilibrium between binary destruction and creation develops in $N$-body simulations, that is the binary fraction does not diminish any more, but it also cannot grow. Thus, the eventual growths by one percent in the binary fractions from 3~Myr to 5~Myr in \BiPoS$\ $are probably due to uncertainties in the fits, rather than to the actual creation of binaries in the $N$-body simulations to which are fitted.

\subsection{The most massive star clusters}
\label{sec:mostmassive}

The IMF of star clusters with the highest masses may have been top-heavy, that is over-abundant in high-mass stars, compared to the canonical IMF (e.g. \citealt{Dabringhausen2009,Dabringhausen2012,Marks2012b}). The mass-range for star clusters where the canonical IMF changes to an increasingly top-heavy IMF has been placed roughly above $10^6 \, {\rm M}_{\odot}$. The slopes of the IMF in \BiPoS$\ $on the other hand are fixed to the ones of the canonical IMF and the user may only change the lower and the upper mass limit for stars. Thus, it may seem questionable how far the results in Section~(\ref{sec:SCexample}) for $M_{\rm ecl}=3\times 10^7 \, {\rm M}_{\odot}$, and especially those for $M_{\rm ecl}=3\times 10^8 \, {\rm M}_{\odot}$, can be trusted.

Moreover, $\Omega_{\rm dyn}^{\rho_{\rm ecl}}(\log_{10}(E_{\rm b}),t)$ in \BiPoS$\ $has been fitted to $N$-body simulations of star clusters in \citet{Marks2011a}. They have embedded cluster masses from $M_{\rm ecl}=10 \, {\rm M}_{\odot}$ to $M_{\rm ecl}=3000 \, {\rm M}_{\odot}$ and initial (embedded) half-light radii from $r_{\rm h}=0.1$ pc to $r_{\rm h}=0.8$ pc. This corresponds to densities within $r_{\rm h}$ between $2.3 \, {\rm M}_{\odot} \, {\rm pc}^{-3}$ and $3.6 \times 10^5 \, {\rm M}_{\odot} \, {\rm pc}^{-3}$, as \BiPoS$\ $calculates the binary survival rates with the densities, rather than relying on combinations of $M_{\rm ecl}$ and $r_{\rm h}$. The most massive star cluster in Section~(\ref{sec:SCexample}) has however $M_{\rm ecl}=3\times 10^8 \, {\rm M}_{\odot}$, and has consequently by default an average density within $r_{\rm h}$ of $1.8 \times 10^7 \, {\rm M}_{\odot} \, {\rm pc}^{-3}$, or about 50 times larger than the maximum density in \citet{Marks2011a}. More broadly speaking, \BiPoS$\ $places no limits on the densities it allows, but the more the densities in \BiPoS$\ $are below or above the densities covered in \citet{Marks2011a}, the more caution is advised. However, $\Omega_{\rm dyn}^{\rho_{\rm ecl}}(\log_{10}(E_{\rm b}),t)$ saturates above a certain $\rho_{\rm ecl}$, so even higher densities make no large differences any more.

\subsection{The meaning of the initial distribution functions}
\label{sec:meaning}

What do the initial distribution functions (IMF, birth binary distribution functions) and values (initial half-mass radius) applied as defaults in \BiPoS$\ $mean? Such distribution functions are needed in order to model stellar populations, because it is not possible to generate realistic populations of stars and binaries on star-cluster- and galaxy-scales from radiation hydrodynamical star-formation simulations.

The IMF is well known not to be observable \citep{Kroupa2018} but it is a mathematical tool needed to synthesize stellar populations across space and time. The mathematical model of the IMF can be constructed from star counts by statistically accounting for all stars formed in one embedded cluster. The IMF shows an insignificant variation with physical conditions in the present-day star-forming activity observed in the Local Group \citep{Bastian2010,Kroupa2013,Offner2014}, but has been found to vary systematically with metallicity and density \citep{Dabringhausen2009,Dabringhausen2012,Marks2012b,Zhang2018,Yan2020}.

Likewise, the birth binary distribution functions are also not observable since no time exists when the fully assembled initial population can be observed because the binary systems evolve upon formation with their formation times not being equal. This can be seen in \citet{Bate2009} and \citet{Bate2012}. In his hydodynamical simulations of the formation of a star cluster, binaries begin to evolve and even to dissolve well before star formation in general has come to an end. But, just as with the IMF, mathematical formulations of the distribution functions can be constructed from the observational data by taking this evolution into account, as detailed in Section~(\ref{sec:theory}). The birth binary distribution functions are, like the IMF, largely invariant for present-day star-formation in the Local Group. But they are expected to vary with extreme conditions, just as the IMF, since both are causally connected \citep{Kroupa2011IAU}. Thus, it is likely that the very metal-poor populations in the Milky Way that formed at a high redshift may have formed with somewhat different distribution functions (IMF, birth binary properties), and that subtle differences may exist between populations formed with different metallicities in the present-day \citep{Liu2019}. 

A simple argument explains why the vast majority of stars should have been be born as binaries: First of all, when a proto-stellar cloud core collapses it will always have too much angular momentum which constitutes a barrier opposing gas infall towards a single centre 
(e.g. \citealt{Goodman1993,Goodwin2007}). If the collapse would in most cases form a stellar system containing more than two stars, then such a system decays on a crossing time scale, which is $10^4-10^5\,$yr. Thus, a triple system would lead to a binary and a single star while a quadruple would decay to a binary and two single stars in two successive decay steps. But low-density (dynamically not significantly processed) star-forming regions show a binary fraction of near unity at an age of about $1\,$Myr. This means that largely only binaries form \citep{Goodwin2005},
with a certain fraction of hierarchical higher-order systems (inner tight binary with a third companion on a wider orbit, say). 

The birth half-mass radius of an embedded cluster, $r_{\rm h}$, has been derived by \citet{Marks2012a} by using the observed binary populations in star clusters to infer their initial density. The derived birth half-mass radii are not observable and represent a mathematical model which is equivalent to the physical reality of the actual formed stellar population. The birth half-mass radii can be understood to represent the instant in time when the entire stellar population appears instantaneously in the deeply embedded cluster.

\subsection{Brown Dwarfs?}
\label{sec:BD}

In \BiPoS, only the binary evolution of stars is treated, but brown dwarfs are missing. If brown dwarfs are understood as objects which have too little mass to ignite hydrogen burning, but are the same as stars in all other respects, including their formation, this poses a potential problem at the lower mass limit of the IMF. The reason is that the IMF in \BiPoS$\ $has $m_{\rm min}=0.08 \, {\rm M}_{\odot}$, that is about the mass required for hydrogen burning, as a lower limit. However, if the objects below $0.08 \, {\rm M}_{\odot}$ are numerous and continue to be star-like, there is no reason why a star above the hydrogen-burning limit should not have a star below the hydrogen-burning limit as a companion. Such binaries, though, are not produced in \BiPoS.

However, using the pairing properties of brown dwarfs and of stars, \citet{Kroupa2003b} concluded that brown dwarfs follow an IMF$_{\rm BD}$, which is related to, but different to that of stars. Following up on this, \citet{Thies2015} normalised the theoretical IMFs from \citet{Chabrier2005} and \citet{Thies2007} to the observed one, such that they produce an equal number of objects in the stellar regime. The shape of the theoretical IMFs and the observed IMF fit in the stellar mass range above $m=0.2 \, {\rm M}_{\odot}$. If they however subtract the theoretical IMFs from the observed one, they find many objects unaccounted for below a mass of $\approx 0.2 \, {\rm M}_{\odot}$. Hence, they conclude that below a mass of $\approx 0.2 \, {\rm M}_{\odot}$, there must be also a process besides direct fragmentation of a gas cloud, which they assume is the process responsible for the production of stars. This could be the production of brown dwarfs from dynamically pre-processed gas \citep{Thies2015}, for example through the fragmentation of circumstellar disks \citep{Stamatellos2009,Thies2010}. The hydrodynamical modelling of \citet{Thies2010} leads indeed to an impressive agreement with observations of the mass distribution of formed brown dwarfs, as well as their binary distributions.

Therefore, \citet{Thies2015} argue that a star-like population and a brown-dwarf-like population of objects \mbox{exist} below stellar masses of $0.2 \, {\rm M}_{\odot}$, while above that mass only the star-like population exists. The star-like population forms objects just like conventional stars, that is stars massive enough to support hydrogen burning. Indeed, those stars are part of the star-like population, but the star-like population continues also to masses with lower masses. The brown-dwarf-like population on the other hand forms its objects differently than the star-like population, but like the star-like population on both sides of the hydrogen burning limit. Note that it is not important here how exactly objects in the two populations form; it suffices that they form differently. Thus, at a mass of $\approx 0.1 \, {\rm M}_{\odot}$, both populations are present. 

There is of course no reason why two different populations should have the same properties. Indeed, \citet{Thies2015} state that the binary fraction at birth of the star-like population is near 100 percent, while the binary fraction at birth of the brown-dwarf-like population is about 20 percent. This makes clear why mingling the two populations in \BiPoS$\ $should be done with extreme care, as \BiPoS$\ $assumes that they all are star-like objects and therefore all have a companion at the beginning. Potential trouble with this can be avoided by only considering stars with a mass larger than $\approx 0.2 \, {\rm M}_{\odot}$, that is by excluding brown-dwarf-like objects. However it makes no sense to extend the range of considered objects in \BiPoS$\ $below $0.08 \, {\rm M}_{\odot}$, because the objects rarely pair at their birth due to their different origin. This is known as the brown dwarf desert (e.g. \citealt{Marcy2000,McCarthy2004}).

In the future, \BiPoS$\ $can be extended to also process the brown dwarf population dynamically.

\subsection{Hardened binaries}
\label{sec:hard}

\BiPoS$\ $currently computes the disruption of binaries in embedded clusters, but the stellar-dynamical processing of a binary population also include the hardening of orbits. A small fraction of hard binaries (those with an orbital velocity larger than the average velocity dispersion in the embedded cluster) will, on average, gain binding energy in an encounter and thus exit the encounter with a smaller semi-major axis and a shorter period and typically a larger eccentricity. The hardening of binary star orbits through stellar encounters can be included into \BiPoS$\ $in the future using statistical rate equations \citep{Heggie1975} and a hardening operator, which quantifies which orbits harden and by how much in a dynamically processed population. The hardened binaries appear in the eccentricity-period diagramme as `forbidden binaries' \citep{Kroupa1995b,Beck2018} before they rapidly circularise and tighten further with possible merging. They thus comprise a potentially powerful tool to study the interiors of stars.

\section{Summary and Conclusion}
\label{sec:summary}

According to \citet{Kroupa1995a} and \citet{Lada2003}, all stars are born in embedded star clusters, although many of these dissolve within a few Myr. A property of stars is that many of them have a companion star, that is that they are binaries. Because the binaries settle into more stable configurations by themselves (pre-main sequence eigenevolution, or internal binary evolution), and are especially in the dense environments of star clusters disturbed by other stars and binaries (stimulated evolution, or external binary evolution), a correct treatment of stellar evolution and stellar dynamics involves a treatment of binary dynamics as well. 

While the existence of binaries is a common feature of all stellar populations, the exact number of binary stars per centre-of-mass systems (that is single stars and multiple stars together) can be quite different from environment to environment \citep{Duchene2013}.

Many stars are born as binaries, but how high is the binary fraction at birth? \citet{Kroupa1995a} and \citet{Goodwin2005} argue that it was extremely high in the beginning and declined afterwards through stellar encounters, and that it can be set for all practical purposes to 100 percent in the beginning. However, chances to observe such a very young population with nearly 100 percent binaries are very small, because the widest binaries dissolve almost instantly to single stars.

The quick dissolution of wide binaries is exactly the point where the computer programme \BiPoS$\ $comes in. Rather than calculating a star cluster from its birth with 100 percent binaries in a very expensive direct N-body simulation (e.g. \textsc{Nbody6}; \citealt{Aarseth1999}), it allows the user to skip the first Myrs of evolution, and set up the star cluster correctly. \BiPoS$\ $solves the problem of the binary evolution within the first Myr based on a stellar-dynamical operator, $\Omega_{\rm dyn}^{\rho_{\rm ecl}}$, introduced in \citet{Kroupa2002} and \citet{Marks2011a}. This operator is basically a function which determines how many binaries survive dynamical evolution in dependency of the binding energy of the binary. It depends on the initial stellar density, $\rho_{\rm ecl}$, within the half-mass radius, $r_{\rm h}$, as well as the time , $t$, for which the dynamical evolution of the binaries proceeds. The latter can be set equal to the time of gas expulsion, which happens after a few Myr and leads to the dramatic expansion, if not the dissolution of the star cluster \citep{Lada1984,Kroupa2001b}. Thus, the environment for efficient external evolution of binaries, or stimulated evolution, is destroyed by gas expulsion. Equivalently to $\rho_{\rm ecl}$, the user may also choose the half-mass radius, $r_{\rm h}$, and the initial embedded cluster mass in stars, $M_{\rm ecl}$, before \BiPoS$\ $starts its calculation.

The operator $\Omega_{\rm dyn}^{\rho_{\rm ecl}}$ has been fitted in \citet{Marks2011a} to numerical N-body simulations done with \textsc{Nbody6} \citep{Aarseth1999}. \citet{Marks2011a} started with star clusters with 100 per cent binaries and evaluated $\Omega_{\rm dyn}^{\rho_{\rm ecl}}$ for 1,~3, and 5~Myr. Thus, these are the values for $t$ available in \BiPoS.

\BiPoS$\ $ calculates the requested orbital parameter(s), such as periods, semimajor-axes, apparent separations, and many more, sorted into bins of binary fraction in intervals of those parameters. The benefit of using \BiPoS$\ $is that it provides an extremely fast and good approximation (solutions within seconds or minutes) to the dynamical processing of a population of binary stars, compared to the exact, but much more time-consuming solution with an N-body programme.

\BiPoS$\ $can also generate galactic field populations, that is stellar populations that are not in star clusters any more. For this, \BiPoS$\ $relies on the assumptions that all stars were born in embedded star clusters, and that these  embedded star clusters follow a power law with a slope of $\beta \approx 2$ \citep{Lada2003}. Thus, \BiPoS$\ $follows essentially the theory of the IGIMF for the stellar population (e.g. \citealt{Kroupa2003,Weidner2005,Kroupa2013,Jerabkova2018,Dabringhausen2019}) when it calculates its binary content. \BiPoS$\ $assumes here that gas expulsion sets in after 3~Myr. It efficiently stops further binary processing in the cluster through its expansion and dissolution, and sheds its evolved population into the field.

In the output of \BiPoS, there are single stars and binaries only, while higher-oder multiples do not appear. They exist in reality though, and their percentage is surprisingly high in massive stars, see for instance \citet{Evans2005}. However, low-mass stars overwhelm high mass stars for in the IMF taken in \BiPoS$\ $(see equation~\ref{eq:imf}), and for all stars, the number of higher-order multiples is small compared to singles and binaries. For instance, \citealt{Duquennoy1991} find 2743 single stars and 2508 binaries in the stellar population of the Galactic Field, but only 23 triples and 14 quadruples. Therefore, higher-order multiples are neglected in \BiPoS, and the success of \BiPoS$\ $with reproducing the observed binary populations lends credence to the initial assumptions on which \BiPoS$\ $is based, namely the shape of the initial period distribution function of the binaries, and the initial binarity of 100 percent.

For the treatise of higher-order multiples, \citet{Hamers2021} recently published the Multiple Stellar Evolution code (\mbox{\textsc{MSE}}). \mbox{\textsc{MSE}} takes a single higher-order multiple system as its input, and evolves it, including its stellar evolution and post-Newtonian terms, besides evolution of its sub-binaries. However, \mbox{\textsc{MSE}} does not, unlike \BiPoS, perform dynamical processing in embedded clusters of an entire birth population of stars. It will be an important task in the future to combine both codes to improve the population synthesis of galactic field populations.

It is also important to study how changing the initial conditions affect the outcome of \BiPoS. In particular, it will be very interesting to investigate degeneracies: for example, a larger birth half-mass radius might be compensated by a larger star formation efficiency, which leads to the same dynamically processed binary-star population. In this case, the initial conditions would be ´dynamically equivalent’ \citep{Belloni2018}.

Note that there are also other, earlier successful attempts to explain binary populations with \BiPoS, although it was never properly introduced until this paper. This concerns the papers by \citet{Marks2011b, Marks2012a}, \citet{Marks2014, Marks2015} and \citet{Marks2017}, even though adapted versions of \textsc{BiPoS} were used for some of them.

There are two caveats regarding the ranges in which \BiPoS$\ $is applied. The first caveat is for star clusters with masses $M_{\rm ecl}>10^6 \, {\rm M}_{\odot}$ in the beginning. This is because GCs with masses $M \apprge 10^6 \, {\rm M}_{\odot}$ today show evidence that their IMF was not canonical, but top-heavy \citep{Dabringhausen2009,Dabringhausen2012,Marks2012b}. The IMF in \BiPoS$\ $is however fixed to the canonical slopes. The second caveat is for stars with masses $m \apprle 0.2 \, {\rm M}_{\odot}$. In this mass range, an additional brown-dwarf-like mass function with different binary properties comes in besides the star-like mass function, which is the single mass function above $m= 0.2 \, {\rm M}_{\odot}$ \citep{Thies2015}. However, \BiPoS$\ $assumes that all stars belong to the star-like population (see \citealt{Marks2015}).

A few examples of how to operate with \BiPoS$\ $are also given in this paper, namely 1.) a comparison between the birth population of binaries with the initial population after internal binary evolution, or pre-main sequence eigenevolution, 2.) a comparison of the binary survival fractions of progenitor models for globular clusters over the range of their embedded stellar masses, 3.) a discussion of the high binary fraction at low apparent separation discovered by \citet{Duchene2018} in the Orion Nebula Cluster, and 4.) a demonstration of the galactic field mode in \BiPoS$\ $with two associations of O- and B-stars.

Future developements of \textsc{BiPoS} might include adding brown dwarfs, hardening of binaries, a variable IMF, and an update of the pre-main sequence eigenevolution, or internal binary evolution \citep{Belloni2017}.

\section*{Acknowledgements}

J\"{o}rg Dabringhausen and Pavel Kroupa acknowledge support from the Grant Agency of the Czech Republic under grant number 20-21855S.

\section*{Data availability}

All data in this paper is either available in the cited literature, or was generated with the computer programme \BiPoS, using the commands detailed in this Paper. \BiPoS$\ $can be downloaded at GitHub under the web adress \texttt{https://github.com/JDabringhausen/BiPoS1}.

\bibliographystyle{mn2e}
\bibliography{bipos}

\label{lastpage}

\end{document}